\documentclass[useAMS,usenatbib]{mn2e}
\usepackage{url}
\usepackage{stmaryrd}
\usepackage[pdftex]{graphicx}
\usepackage{natbib}
\usepackage[backref,breaklinks,colorlinks,citecolor=blue]{hyperref}
\usepackage[T1]{fontenc}
\usepackage{aecompl}
\usepackage[all]{hypcap}

\usepackage{color}

\definecolor{darkblue}{rgb}{0.0, 0.0, 0.5}
\hypersetup{colorlinks=true,citecolor=darkblue,linkcolor=darkblue}

\def\mjup{M$_{Jup}$}
\def\mj{M_{Jup}}
\def\msun{M$_{\odot}$}

\def\gtrsim{\ {\raise-.5ex\hbox{$\buildrel>\over\sim$}}\ }
\def\ltsim{\ {\raise-.5ex\hbox{$\buildrel<\over\sim$}}\ } 

\begin{document}

\title[J1407 companion mass limits]{Mass and period limits on the ringed companion transiting
  the young star J1407}

\author[M.A. Kenworthy et al.]{
M.~A. Kenworthy$^{1}$\thanks{E-mail: kenworthy@strw.leidenuniv.nl},
S. Lacour$^{2}$,
A. Kraus$^{3}$,
A.~H.~M.~J. Triaud$^{4,5}$
\footnote{Fellow of the Swiss National Science Foundation},
\newauthor
E.~E. Mamajek$^{6}$,
E.~L. Scott$^{6}$,
D. S\'egransan$^{5}$,
M. Ireland$^{7}$,
F.-J. Hambsch$^{8}$,
\newauthor
D.~E. Reichart$^{9}$,
J.~B. Haislip$^{9}$,
A.~P. LaCluyze$^{9}$,
J.~P. Moore$^{9}$,
N.~R. Frank$^{9}$
\\
$^{1}$ Leiden Observatory, Leiden University, P.O. Box 9513, 2300 RA Leiden, the Netherlands\\
$^{2}$ LESIA, CNRS/UMR-8109, Observatoire de Paris, UPMC, Universit\'{e} Paris Diderot, 5 place Jules Janssen, 92195 Meudon, France \\
$^{3}$ Institute for Astronomy, University of Hawaii, 2680 Woodlawn Drive, Honolulu, HI 96822, USA \\
$^{4}$ Kavli Institute for Astrophysics \& Space Research, Massachusetts Institute of Technology, Cambridge, MA, USA\\
$^{5}$ Observatoire Astronomique de l'Universit\'{e} de Gen\`{e}ve, Chemin des Maillettes, 51, CH-1290 Sauverny, Switzerland\\
$^{6}$ Department of Physics and Astronomy, University of Rochester, Rochester, NY 14627-0171, USA \\
$^{7}$ Research School of Astronomy and Astrophysics, Mt Stromlo Observatory, Cotter Rd, Weston ACT 2611, Australia \\
$^{8}$ Center for Backyard Astrophysics (Antwerp), American Association
of Variable Star Observers (AAVSO),\\Vereniging Voor Sterrenkunde (VVS), Andromeda Observatory, Oude Bleken 12, 2400 Mol, Belgium\\
$^{9}$ Department of Physics and Astronomy, University of North Carolina
at Chapel Hill, Campus Box 3255, Chapel Hill, NC 27599, USA}
\date{Accepted for publication in MNRAS on 01 Oct 2014}

\pagerange{\pageref{firstpage}--\pageref{lastpage}} \pubyear{2014}

\maketitle

\label{firstpage}

\begin{abstract}

The young ($\sim 16$ Myr) pre-main-sequence star in Sco--Cen 1SWASP
J140747.93-394542.6, hereafter referred to as J1407, underwent a deep
eclipse in 2007 April, bracketed by several shallower eclipses in the
surrounding 54 d.
This has been interpreted as the first detection of an eclipsing ring
system circling a substellar object (dubbed J1407b).
We report on a search for this companion with Sparse Aperture Mask
imaging and direct imaging with both the UT4 VLT and Keck telescopes.
Radial velocity measurements of J1407 provide additional constraints on
J1407b and on short period companions to the central star.
Follow-up photometric monitoring using the PROMPT-4 and ROAD
observatories during 2012-2014 has not yielded any additional eclipses. 
Large regions of mass-period space are ruled out for the companion.
For circular orbits the companion period is constrained to the range
3.5-13.8 yr ($a$ $\simeq$ 2.2-5.6 au), and stellar masses
($>$80\,\mjup) are ruled out at 3$\sigma$ significance over these
periods.
The complex ring system appears to occupy more than 0.15 of its
Hill radius, much larger than its Roche radius and suggesting a ring
structure in transition.
Further, we demonstrate that the radial velocity of J1407 is consistent
with membership in the Upper Cen--Lup subgroup of the Sco--Cen
association, and constraints on the rotation period and projected
rotational velocity of J1407 are consistent with a stellar inclination
of $i_{\star}$ $\simeq$ 68$^{\circ}$\,$\pm$\,10$^{\circ}$.

\end{abstract}

\begin{keywords}
planets and satellites: formation -- planets and satellites: rings --
binaries: eclipsing--planetary systems--stars: individual: 1SWASP J140747.93-394542.6 (ASAS J140748-3945.7).
\end{keywords}

\section{\label{s:intro}Introduction}

An otherwise unremarkable pre-main sequence star in the nearby Sco--Cen
OB Association, 1SWASP J140747.93-394542.6 (hereafter J1407) suddenly
exhibited an extremely long, deep, and complex eclipse in mid-2007,
lasting 54 days and achieving a maximum depth of $>$3 magnitudes
\citep{Mamajek12}. This eclipse morphology is markedly different when
compared with a typical eclipsing binary or exoplanet, but is very
reminiscent of the light curves for stars such as $\epsilon$ Aurigae
\citep{Guinan02,Kloppenborg10,Chadima11}, KH 15D \citep{Winn06b}, EE
Cep \citep{Mikolajewski99,Graczyk03,Mikolajewski05} that are
periodically occulted by extended objects. Given its youth, J1407 is
most likely being eclipsed by a low-mass object hosting a disk with
significant substructure composed of thin dust debris belts, or
`rings'. There is no evidence of flux from this companion,
indicating that it is a very low mass star, brown dwarf, or perhaps
even a gas giant planet \citep{Mamajek12}.

J1407 was identified as a probable young star by virtue of its X-ray
emission, a proper motion consistent with the motion of Sco--Cen, and
its CMD position on the Sco--Cen stellar sequence
\citep{Mamajek12,Pecaut13}.  Spectroscopic follow-up using the CTIO
1.5m telescope confirmed the object to be a young, Li-rich K5 star,
lacking any accretion indicators \citep{Mamajek12} or infrared excess
at short \citep[2MASS; ][]{Skrutskie06} and long wavelengths
\citep[WISE photometry; ][]{Wright10}.  The star therefore appears to
be a weak-lined T Tauri star, not hosting its own circumstellar disk,
and is a young ($\sim$16 Myr) solar analogue ($M\sim$0.9 $M_{\odot}$).

This star was observed by the Super Wide Angle Search for Planets
(SuperWASP) survey \citep{Pollacco06,Butters10} in 2007 as part of
their wide-field search for transiting exoplanets.  Simultaneously and
independently, the field was also observed at much lower cadence by
the All Sky Automated Survey \citep[ASAS; ][]{Pojmanski02}. Both
surveys revealed a remarkably long, deep, and complex eclipse event
centred on UT 2007 April 29. At least five multiday dimming events of
$>$0.5 mag were identified, with a $>$3.3 mag deep eclipse bracketed
by two pairs of $\sim$1 mag eclipses symmetrically occurring $\pm 12$
d and $\pm 26$ d before and after. Hence, significant dimming of
the star was taking place on and off over at least a $\sim$54 d
period in 2007, and a strong $>$3 mag dimming event occurring over a
$\sim$12 d span. \citet{Mamajek12} hypothesized that a complex ring
system circling an unseen lower mass companion passed in front of
J1407 over the span of several weeks.

The WASP pipeline is optimized for identifying transiting planets,
which typically produce trough-like eclipses of 1\% depth lasting a
few hours. In the case of J1407, it was located in the corner of the
field of view of three of the WASP-South cameras, and the photometry
from different cameras showed systematic relative offsets of up to 0.3
mag. A custom reduction pipeline was created for J1407 and
these systematic effects, along with the rotational variability, were
removed to produce a cleaned light curve \citep{vanWerkhoven14}. 

For orbital periods of 20--100 yr ($a > 7$\, au), the projected
separation is in the range of $\sim$60 mas. Given the likely faintness
of the companion, it falls well below the angular resolving limits of
traditional adaptive optics imaging on 8-10m class telescopes.  Sparse
aperture masking (SAM) or Non-Redundant Masking (NRM) delivers an
increase in resolution within the Airy disk diffraction limit
\citep[e.g.  ][]{Nakajima89,Tuthill06}, and is now a well established
means of achieving the full diffraction limit of a single telescope
\citep[e.g.  ][]{Ireland08,Kraus09,Lacour11}.

Predicting when the next eclipse will happen is paramount to planning
an extensive and detailed observing campaign. This object offers an
unprecedented opportunity to spatially and spectrally resolve a disk
with ringlike structure orbiting a likely substellar object at age
$\sim 16$\ Myr. In this paper we report on our search for the proposed
companion J1407b. In Section \ref{obsred} we describe the observations
and data reduction. We do not detect any companion to J1407 with
interferometric or direct imaging, and high spectral resolution radial
velocity (RV) measurements show no companion in the system.  Using these
constraints, we provide a simple model for circular and elliptical orbital solutions
(Section \ref{an}) and show which orbital periods can be ruled out. We
discuss the consequences of these observations for the J1407 system in
Section \ref{co}.

\section{Observations and data reduction}
\label{obsred}

J1407 has been observed with several different instruments, all
providing different constraints on companions in the system. In the
first three sections we present interferometric and direct imaging
observations from the VLT and Keck. The next two sections detail
radial velocity (RV) measurements from MIKE on the Magellan
telescope and CORALIE on the Euler telescope. The last sections detail
photometric monitoring with the PROMPT-4 telescope and the Remote
Observatory Atacama Desert (ROAD) observatory.

\subsection{SAM at the VLT}

A Sparse Aperture Mask (SAM) takes the filled single dish telescope
pupil and selects several smaller apertures from it. The locations of
these apertures are chosen so that all pairwise combinations form a set
of unique baselines. The resultant point spread function (PSF) in the focal
plane of the science camera then encodes spatial information about the
science target that can be reconstructed to form a model of the source.
Slow changing aberrations within the telescope and science instrument
can be calibrated out by interleaving science observations with those of
a similar brightness reference source. The outer size is set by the
shortest baseline between two apertures and an inner size set by the
interferometric limit $(\lambda/2D)$. Complete baseline coverage is
obtained thanks to the rotation of the target along the parallactic
angle during the observation. Although there is a penalty in reduced
throughput, SAM imaging can reach imaging scales that are not easily
attainable with classical adaptive optics assisted direct imaging.  This
interferometric method is also referred to as NRM and this term can be interchangeably used with SAM imaging. We use
SAM for the rest of this paper.
For both Keck and VLT, maximizing throughput using the sparsest mask with the largest
apertures is almost always better for detecting point source companions
than maximising baselines (using a denser mask with smaller apertures).
As the target star is relatively faint for SAM imaging and the secondary
companion is expected to be a point source, choosing as few baselines as
possible helps optimize signal to noise (S/N).

Several SAM modes with different aperture configurations are
implemented with the NaCo system at the VLT \citep{Lacour11}.
Observations were taken with VLT NaCo on UT 2013 March 27 and 28.  We
use the `7-hole' mask for maximum throughput. The conditions for
both nights were photometric with no cloud cover. The VLT seeing
monitor reported 0.5 to 0.9 arcsecond seeing on both nights, with
coherence times of $4$ ms over the whole observing run (see Table
\ref{obslog}).  J1407 has a visual magnitude $V=12.31\pm 0.03$
\citep{Mamajek12} and $K_s=9.26\pm 0.02$ \citep[2MASS; ][]{Cutri03}.
The properties of the star J1407 are detailed in
\citet{vanWerkhoven14} and we use them for the rest of this paper --
the distance is taken to be $133\pm 13 $pc with an age of $16$ Myr.
We use the 90\% IR dichroic and the infrared wavefront sensor (WFS) so that the target
star is used as a natural guide star for the adaptive optics
system.

The telescope optics and instrument have a time-varying optical
aberration associated with them, and so SAM observations of the
science (SCI) target require known single star observations (CAL
observations) to be interspersed between them. Moving between the
science and calibration stars incurs an observing overhead, and so a
balance must be found between remaining on the science target and the
change in telescope aberrations that cannot be removed by sparse
calibration observations. The time between SCI and CAL observations is
determined from previous observing experience to be of the order of a
few minutes.  Observations are alternated between J1407 (SCI) and two
calibration stars chosen from previous SAM observing runs 2MASS
J14074401-3941140 and 2MASS J14072080-3948550. Even though these stars
were used as calibrators in other previous observations by us, there
is always the chance that a calibrator is a long period binary and
that the data is rendered uncalibratable. To mitigate this, two
calibration stars with similar magnitudes and colours to the science
target are chosen.
Between the first and second night it was discovered that the first
night observations were photon noise limited and not calibration noise
limited, and so the dwell time between SCI and CAL observations was
increased from 50--300 s.
Data for J1407 were taken in an eight-point dither pattern on the
first night, and subsequently in a four-point dither pattern on the
second night, resulting in an increase in S/N ratio.
The observing parameters of both J1407 and
the calibration stars are detailed in Table \ref{starprop}.

\begin{table*}
\centering
    \caption{Observing log for the SAM imaging. All data are taken with
the $K_s$ filter and `7 hole' SAM. \label{obslog}}
    \begin{tabular}{lcccccccc}
\hline
Star & UT start & DIT & ${\rm N_{DIT}}$ & Pointings & Seeing & $\tau_0$  \\
 &          & (s) &      & offsets  & (arcsec) &  (ms) \\
\hline
          J1407 & 2013-03-27 05:01:41 & 10 &  5 & 8 & 1.16 & 2.53 \\
    2M1407-3941 & 2013-03-27 05:15:51 &  5 & 10 & 8 & 0.82 & 3.44 \\
          J1407 & 2013-03-27 05:29:06 & 10 &  5 & 8 & 0.74 & 3.86 \\
    2M1407-3948 & 2013-03-27 05:44:38 &  7 &  7 & 8 & 0.73 & 3.83 \\
          J1407 & 2013-03-27 05:56:37 & 10 &  5 & 8 & 0.66 & 4.21 \\
    2M1407-3941 & 2013-03-27 06:06:54 &  5 & 10 & 8 & 0.76 & 3.63 \\
          J1407 & 2013-03-27 06:18:25 & 10 &  5 & 8 & 0.80 & 3.48 \\
    2M1407-3948 & 2013-03-27 06:29:28 &  7 &  7 & 8 & 0.96 & 2.82 \\
          J1407 & 2013-03-27 06:42:50 & 10 &  5 & 8 & 0.86 & 3.29 \\
    2M1407-3941 & 2013-03-27 06:55:34 &  5 & 10 & 8 & 0.70 & 3.99 \\
          J1407 & 2013-03-27 07:06:17 & 10 &  5 & 8 & 0.75 & 3.90 \\
    2M1407-3948 & 2013-03-27 07:17:51 &  7 &  7 & 8 & 0.74 & 3.84 \\
          J1407 & 2013-03-27 07:28:52 & 10 &  5 & 8 & 0.71 & 3.99 \\
    2M1407-3941 & 2013-03-27 07:39:26 &  5 & 10 & 8 & 0.61 & 4.65 \\
          J1407 & 2013-03-27 07:50:22 & 10 &  5 & 8 & 0.62 & 4.59 \\
    2M1407-3948 & 2013-03-27 08:01:02 &  7 &  7 & 8 & 0.61 & 4.66 \\
          J1407 & 2013-03-27 08:11:37 & 10 &  5 & 8 & 0.64 & 4.55 \\
    2M1407-3941 & 2013-03-27 08:22:56 &  5 & 10 & 8 & 0.65 & 4.55 \\
          J1407 & 2013-03-27 08:40:11 & 12 &  4 & 8 & 1.03 & 2.86 \\
    2M1407-3948 & 2013-03-27 08:51:28 &  7 &  7 & 8 & 0.84 & 3.63 \\
\hline
          J1407 & 2013-03-28 04:25:45 & 10 & 30 & 4 & 0.70 & 3.25 \\
    2M1407-3941 & 2013-03-28 04:50:02 & 10 & 30 & 4 & 0.85 & 2.94 \\
          J1407 & 2013-03-28 05:13:54 & 10 & 30 & 4 & 0.69 & 3.59 \\
    2M1407-3941 & 2013-03-28 05:38:07 & 10 & 30 & 4 & 0.64 & 3.74 \\
          J1407 & 2013-03-28 06:14:46 & 10 & 30 & 4 & 0.65 & 3.62 \\
    2M1407-3941 & 2013-03-28 06:39:22 & 10 & 15 & 4 & 0.77 & 3.19 \\
          J1407 & 2013-03-28 06:52:45 & 10 & 30 & 4 & 0.73 & 2.81 \\
    2M1407-3948 & 2013-03-28 07:18:27 & 10 & 21 & 4 & 0.67 & 3.50 \\
          J1407 & 2013-03-28 07:36:47 & 10 & 30 & 4 & 0.58 & 3.78 \\
    2M1407-3941 & 2013-03-28 08:01:13 & 10 & 15 & 4 & 0.68 & 3.83 \\
          J1407 & 2013-03-28 08:15:36 & 10 & 30 & 4 & 0.81 & 2.87 \\
    2M1407-3948 & 2013-03-28 08:40:42 & 10 & 21 & 4 & 0.69 & 3.18 \\
          J1407 & 2013-03-28 09:00:14 & 10 & 30 & 4 & 0.60 & 3.88 \\
    2M1407-3941 & 2013-03-28 09:24:51 & 10 & 15 & 4 & 0.58 & 3.82 \\
          J1407 & 2013-03-28 09:37:56 & 10 & 30 & 4 & 0.62 & 3.55 \\
    2M1407-3948 & 2013-03-28 10:01:58 & 10 & 21 & 4 & 0.91 & 2.41 \\
\hline
\end{tabular}
\end{table*}

\begin{table*}
 \centering
    \caption{Properties of stars observed\label{starprop}}
    \begin{tabular}{lcccccc}
    \hline
Object &  $K_s$ & Total integration time\\
       &  (mag) & (s)\\
\hline
2MASS J14074792-3945427  (SCI) & $9.257\pm0.020$ & 13584 \\
2MASS J14074401-3941140  (CAL) & $8.345\pm0.026$ & 6200 \\
2MASS J14072080-3948550  (CAL) & $8.874\pm0.021$ & 4480 \\
\hline
\end{tabular}
\end{table*}

The VLT data were reduced by fitting the diffraction pattern on each
one of the detector images. Each image of a data cube consists in a
number of fringes of given spatial frequency (proportional to the
baseline vector between each pair of holes), and given phase. The
procedure is explained in more details in \citet{Lacour11}, but
roughly, the phases and amplitudes are used to derive the uncalibrated
visibilities, which are in turn used to compute the bispectrum. The
data reduction is done similarly on the Keck data, but with a notable
difference that the visibility measurements are obtained by fast
Fourier transform. The bispectrum is then co-added over all the frames
of one data cube, of which the argument is taken as the closure
phase. The final step for the J1407 data is to correct for biases in
the closure-phase measurement, by subtracting the closure phase values
of the two calibrators closest in time.

The sensitivity map is obtained by the mean of calculating the $\chi^2$
between the data and a binary model. We used the binary model from
\citet{LeBouquin12} to compute synthetic closure phases:
\begin{equation} 
CP={\rm arg}[(1+\rho e^{i \alpha_{ij}})(1+\rho
e^{i \alpha_{jk}})(1+\rho e^{i \alpha_{ki}})] 
\end{equation}
where $\rho$ is the flux ratio between the two stars, and $\alpha_{ij}$ is a function of the baseline vector between to
two holes $i$ and $j$, but also a function of the separation between
the primary and the secondary stars ($\Delta$): \begin{equation}
  \alpha_{ij}=2\pi\bf B_{ij} \cdot \bf \Delta / \lambda
  \, \end{equation} The $\chi^2$ is then computed over a three
dimensional grid ($\rho$,$\bf \Delta$), and normalized so the $\chi^2$
of a non detection ($\Delta=0$) is equal to the number of closure
phase measurements. Finally, for any separation $\bf \Delta$, the
sensitivity limit is obtained by taking the $\rho$ value where the
$\chi^2$ is equal the number of closure phases plus 25.  Note that
this $\chi^2 + 25$ limit corresponds to a 3.3$\sigma$ and not a
5$\sigma$ sensitivity limit because of the covariance of the closure
phases (for a 7 hole mask, there are 15 independent quantities
instead of 35). We conservatively adopt all subsequent sensitivity
maps as being $3\sigma$ point source detection limited.

\begin{figure*}
\centering
\includegraphics[angle=270,width=\textwidth]{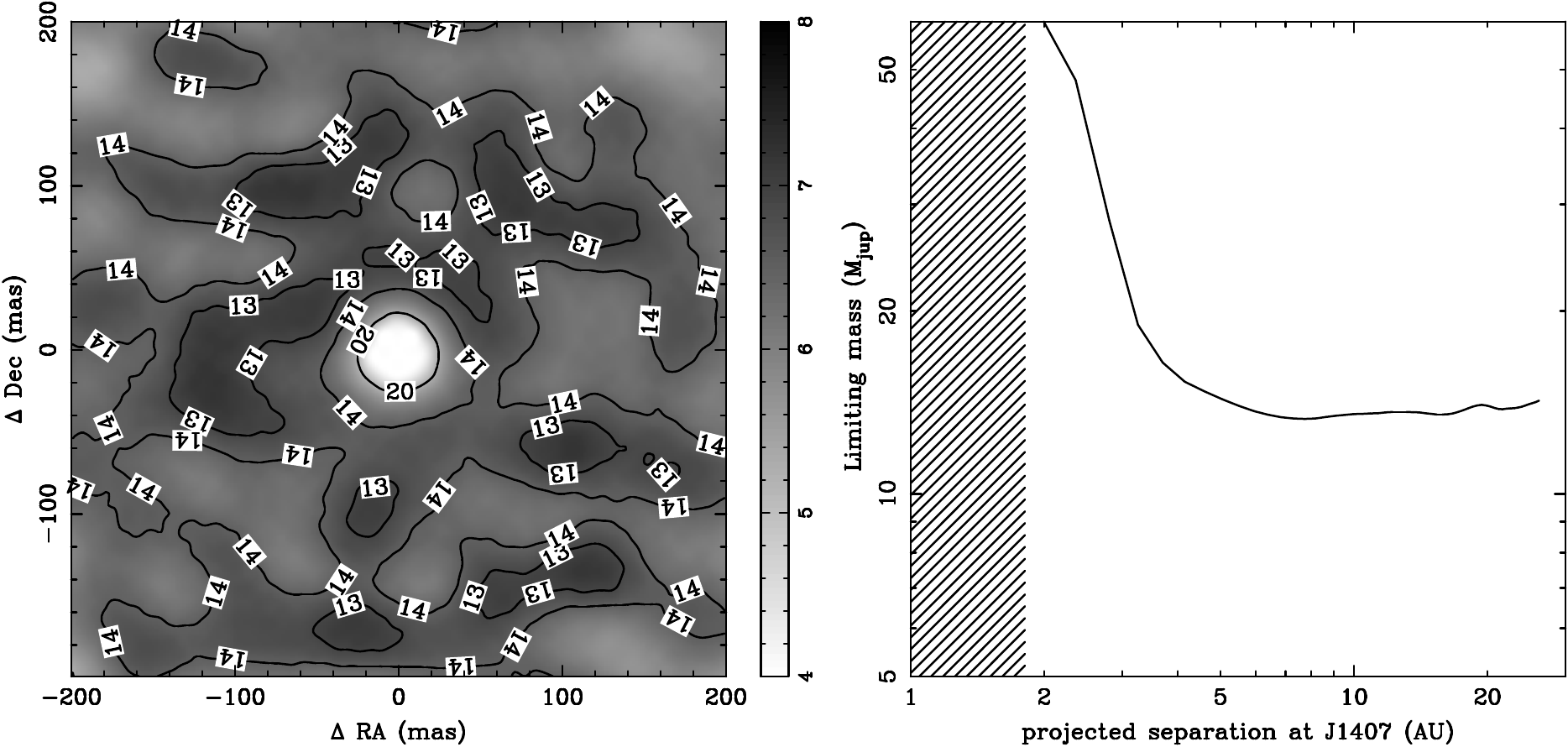}

\caption{Sensitivity map of J1407 from SAM imaging at the VLT. Left-hand panel: the
  star is at the centre of the image, the grey scale and side bar
  indicate the sensitivity in delta $K_s$ magnitudes for a $3\sigma$
  point source detection. The contours indicate the associated upper
  mass limits in units of Jupiter masses (\mjup) estimated from the $K_s$
  photometry and models from \citet{Allard12}.  North is up and east
  to the left.
Right-hand panel: azimuthally averaged sensitivity map. The hashed region indicates circular orbital periods of
  $P < 850$ d that are ruled out by \citet{Mamajek12}. 
\label{samsense}}

\end{figure*}

The results of the observations are shown in Fig. \ref{samsense}. We
do not detect any point sources within the 400 mas region surrounding
J1407. Due to the combination of observing conditions and sky
rotation, the sensitivity for a $3\sigma$ detection varies by 1
mag at a given radius. The sensitivity map is shown in $\Delta
K_s$ magnitudes as a grey-scale image. We use BT-SETTL models without
irradiation and Caffau solar abundances \citep{Caffau11,Allard12} to
estimate the limiting companion mass from the delta $K_s$ magnitude,
the distance to the star and its apparent magnitude. The resultant
mass limits are shown as the contour map in Fig. \ref{samsense} in
units of Jupiter mass (\mjup). It can be seen that masses below 13
\mjup\, are reached in several regions around J1407, rising to 20
\,\mjup\, for radii of 28 mas (3 AU) and smaller.

\subsection{SAM at Keck}

We observed J1407 on UT 2012 April 4 using the Keck-II 10m telescope
with natural guide star adaptive optics. All observations were
conducted with the facility AO imager, NIRC2, which has aperture masks
installed in the cold filter wheel near the pupil stop. We used a
9-hole aperture mask, which yields 28 independent baseline
triangles about which closure phases are measured. All SAM
observations operate in a subarray mode of the narrow camera, and we
conducted our observations using the broad-band $K'$ filter. The
observing sequence consisted of one set of 21 integrations of 20s for
J1407, preceded and followed by similar observations of independent
calibrator stars. We summarize these observations in Table
\ref{obskeck}.

\begin{table*}
 \centering
    \caption{Keck-II observations\label{obskeck}}
    \begin{tabular}{lll}
    \hline
Object & Num. frames & Notes \\
\hline
PDS 70  & 21 & Calibrator \\
J1407 & 21 &  \\
14213051 & 13 & Binary $(\rho = 1 {\rm arcsec},\Delta K=0)$ \\
14085608 & 21 & Binary $(\rho = 230 {\rm mas},\Delta K=2)$   \\
MML 40 & 12 &  Calibrator \\
\hline
\end{tabular}
\end{table*}

\begin{figure*}
\centering
\includegraphics[angle=270,width=\textwidth]{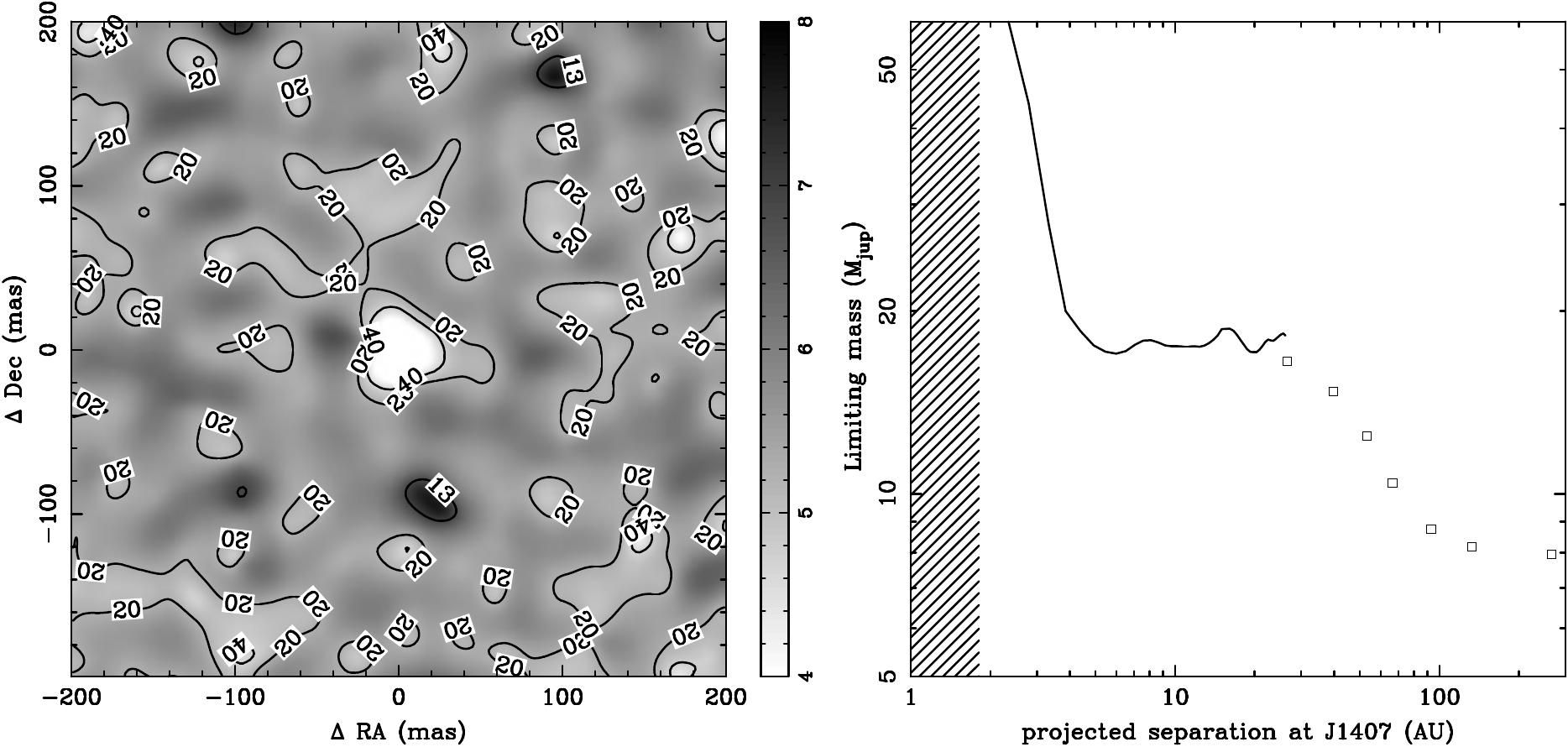}

\caption{Sensitivity map of J1407 from SAM imaging at Keck. {\it Left
panel:} the
  star is at the centre of the image, the grey scale and side bar
  indicate the sensitivity in delta $K_s$ magnitudes for a $3\sigma$
  point source detection. The contours indicate the associated upper
  mass limits in units of Jupiter masses (\mjup) estimated from the $K_s$
  photometry and models from \citet{Allard12}.  North is up and East
  to the left.
{\it Right panel:} Azimuthally averaged sensitivity map. The hashed region indicates circular orbital periods of
  $P < 850$ days that are ruled out by \citet{Mamajek12}. The open
squares are the sensitivity limits from direct imaging observations with
Keck.
\label{fig:kecksam}}

\end{figure*}

The analysis for the SAM data was identical to that used in previous
papers \citep[e.g.,][]{Ireland08a,Kraus08,Kraus11}, combined with the
new calibration technique described in \citet{Kraus12}. To summarize,
the images were flat fielded and bad pixels were removed by
interpolating between neighbouring pixels. The image was then
multiplied by a super-Gaussian window function of the form
$\exp(-ar^4)$, with $r$ the radius in pixels from the centre of the
interferogram. A two-dimensional Fourier transform was then made of
each exposure in a visit, and this Fourier transform was point sampled
at the positions corresponding to the baseline vectors in the aperture
mask. For each visit we then computed the vector of mean uncalibrated
closure phases and the standard error of the mean. Finally, we
calibrated the closure phases for each visit using an optimal linear
combination of the calibrators observed in the same sequence of
events. Our analysis found no statistically significant signal in the
calibrated closure phases for J1407, and hence that it is single to
within the detection limits of the observations. Using the same
procedures as in our previous SAM work mentioned above (i.e., a Monte
Carlo method that simulates random closure phase data sets of a point
source with closure-phase errors and covariances that match those of
the real data), we found the contrast limits summarized in Fig.
\ref{fig:kecksam}.

\subsection{Direct Imaging with Keck}

J1407 was also observed with standard AO imaging immediately preceding
the SAM observations, obtaining six integrations of 10s in the $K'$
filter. These observations used the narrow camera in full-frame mode
(FOV$=$10 arcsec) and were obtained in a diagonal two-point dither
pattern.  We analysed each frame of imaging data using two
complementary methods of PSF subtraction. For wide separations ($\rho
> 600$ mas), where read noise dominated the error budget for companion
detection, we subtracted an azimuthally smoothed median PSF. For small
separations ($\rho \le 600$ mas), where speckles from the primary star
dominated the error budget, we searched the library of all single
stars observed that night to find the best-fitting empirical
template. We determined the best-fitting comparison star using the
pixels at projected separations of $150 < \rho < 300$ mas, scaling
each comparison star to the same total flux and then computing the
reduced $\chi^2$ of the fit. We then subtracted the best-fitting
template star for each frame (implicitly a LOCI subtraction with $n=1$
and globally optimized). Finally, we stacked the residual frames and
measured the standard deviation of fluxes measured through 4 pixel
apertures in radial bins. We found no candidate companions in the
stacked residuals that were more significant than 6$\sigma$, and hence
we report a null detection with the 6$\sigma$ detection limits shown
in Fig. \ref{fig:kecksam}. The direct imaging limits at large radii
reach down to 8\,\mjup, rising to 18\,\mjup\, at 26 au.

\subsection{RV Measurements with Magellan/MIKE}

We obtained high-resolution spectra for J1407 on three consecutive
nights starting on UT 2013 Feb 02 using the Magellan Inamori Kyocera
Echelle (MIKE) optical echelle spectrograph on the Clay telescope at
Magellan Observatory. We used the 0.7 arcsec slit, which yields
spectral resolution of $R=35,000$ across a range of $\lambda =
3350$-9500\AA.  The pixel scale oversamples the resolution with the
0.7 arcsec slit, so we observed with two times binning in the
spatial and spectral directions to reduce readout overheads. The
details of this observing campaign (such as selection of standards)
can be found in \citet{Kraus14}.

We reduced the raw spectra using the {\tt CarPy} pipeline (Kelson
2003)\footnote{\url{http://code.obs.carnegiescience.edu/mike}}. In
order to correct for residual wavelength errors (due to flexure and
uneven slit illumination), we then cross-correlated the 7600\AA\,
telluric $A$ band for each spectrum against a well-exposed spectrum of a
telluric standard, solving for the shift that places each spectrum
into a common wavelength system defined by the atmosphere.  Finally,
we used the procedures described in \citet{Kraus14} to estimate the
spectral type and measure $v_{rad}$, $v \sin(i_{\star})$, EW$[{\rm H}\alpha]$, and
$EW[{\rm Li}_{6708}]$.  We adopt $v \sin(i_{\star})$ = 14.6\,$\pm$\,0.4
km\,s$^{-1}$, which we use later for a calculation of the minimum
radius of the star J1407.  In Table \ref{tab:mike}, we list the epochs
and exposure times for our MIKE observations, the $S/N$ measurement
for each spectrum at 6600\AA, and the values measured from each
spectrum.

\begin{table*}
\centering
\caption{MIKE Observations of J1407\label{tab:mike}}
\begin{tabular}{cccccccc}
\hline
Epoch & $t_{int}$ & SNR & $v_{rad}$ & $v\sin(i_\star)$ & SpT & $EW[H\alpha]$ &
$EW[Li_{6708}]$ \\
(HJD) & (sec) & @6600\AA & (km/s) & (km/s) & (km/s) & (\AA) & (\AA) \\
\hline
56325.81 & 360 & 113 & 5.9 $\pm$ 0.3 & 14.9 $\pm$ 0.4 & K4.8 & -0.36 & 0.451 \\
56326.84 & 720 & 164 & 5.5 $\pm$ 0.3 & 14.4 $\pm$ 0.4 & K4.8 & -0.23 & 0.441 \\
56327.87 & 360 & 119 & 6.4 $\pm$ 0.3 & 14.6 $\pm$ 0.5 & K4.8 & -0.23 & 0.454 \\
\hline
\end{tabular}
\end{table*}

\subsection{RV Measurements with CORALIE}

J1407 was observed using the CORALIE spectrograph, mounted on to the
1.2m Euler Telescope, installed at ESO's La Silla Observatory in Chile.
This is a fibre-fed, thermally-stabilized, \'echelle spectrograph with a
resolution $\sim 60,000$. After a major upgrade in 2007 that improved
its efficiency without comprising stability and precision, it has been
routinely used in determining which of the candidate systems detected by
SuperWASP are bona fide planets. Over seven years, it has detected over
100 transiting exoplanets \citep{Wilson08b,Triaud11,Hellier14}.  J1407's
brightness falls well within CORALIE's typical range of observation.

The spectra have been reduced using CORALIE's Data Reduction Software,
which was built alongside those employed on HARPS, HARPS-South, and
SOPHIE and that have been shown to yield remarkable precision and
accuracy \citep{Dumusque12,Molaro13,Pepe13}.  The spectra are
wavelength calibrated using a Thorium--Argon calibration lamp
\citep{Lovis07}. Each exposure is a simultaneous observation of the
star and of a Th--Ar lamp.  Calibration exposures throughout the night
ensure that the instrumental drifts, mostly due to variations in
pressure (for instance the solar-induced atmospheric-tides), are
corrected. The stellar spectra are cross-correlated using a weighted
numerical K5-spectral mask, following the methods of
\citet{Baranne96}. Radial velocities are computed from fitting a
Gaussian profile on to the resulting cross-correlation function. Other
parameters such as the span of the bisector slope and the full width at
half-maximum (FWHM) of the cross-correlation function were measured as
well. They provide good diagnostics for stellar activity
(e.g. \citealt{Queloz01,Huelamo08}).

\begin{figure}
\centering
\includegraphics[angle=0,width=\columnwidth]{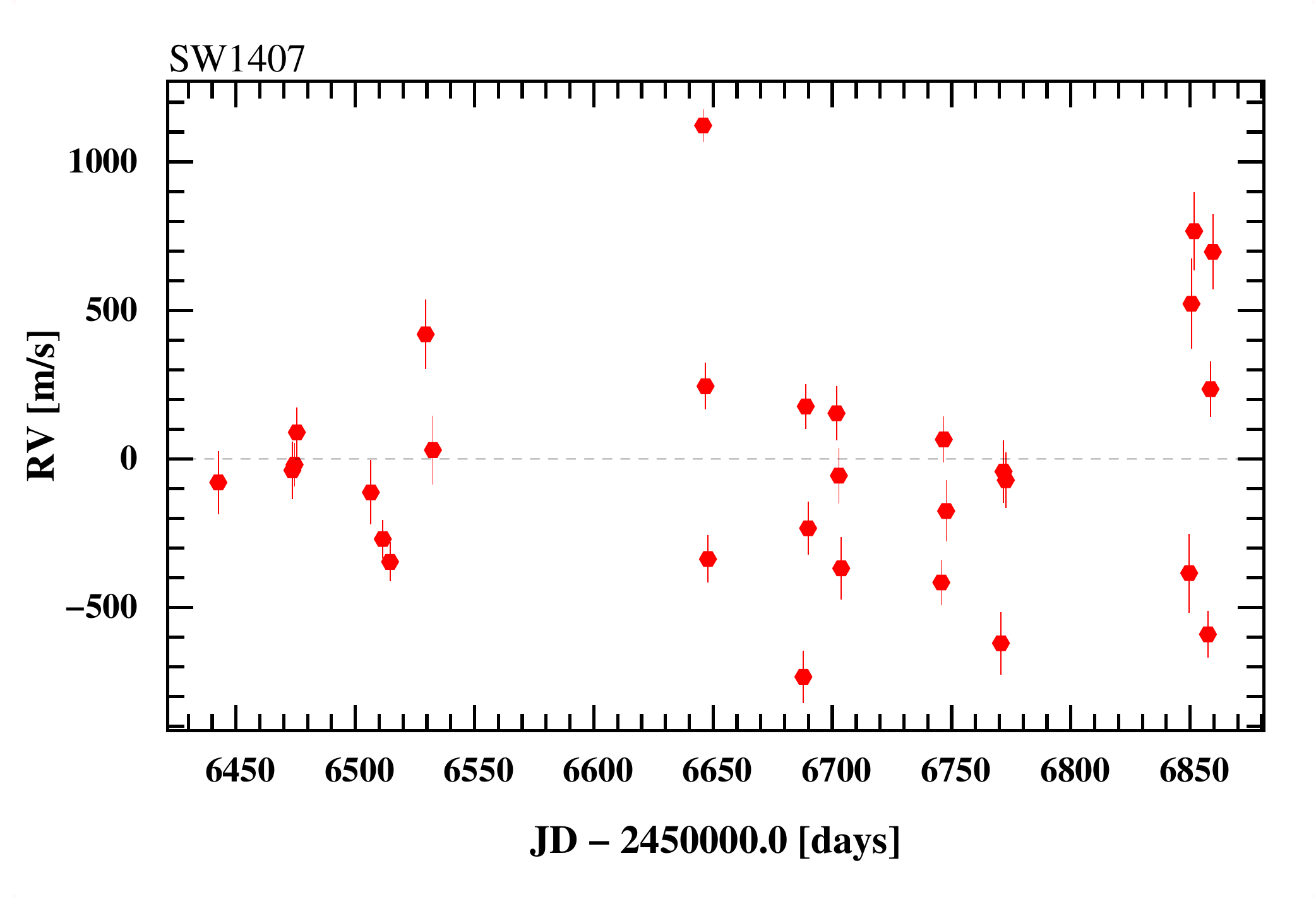}
\caption{CORALIE measurements of the RV of J1407. The mean
systemic velocity $\gamma = 6.91 \pm 0.04$ km s$^{-1}$
has been subtracted off the data. \label{fig:cor1}
}
\end{figure}

\begin{figure}
\centering
\includegraphics[angle=0,width=\columnwidth]{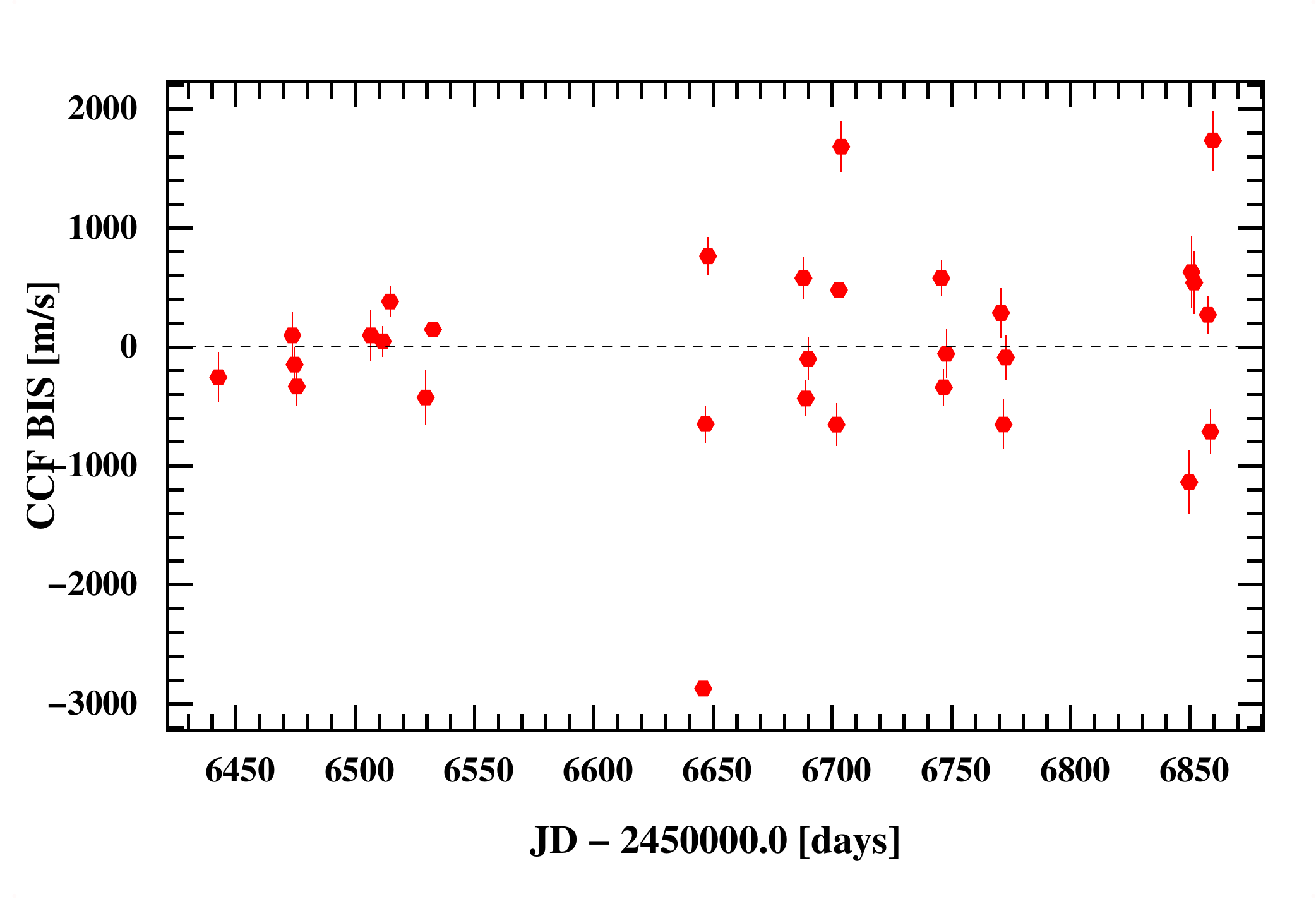}

\caption{CORALIE measurements of the span in the slope of the
  bisector.  The scatter has increased during the second season of
  observations. This is thought to be due to an increase in activity of
the star. \label{fig:cor2} }

\end{figure}

\begin{figure}
\centering
\includegraphics[angle=0,width=\columnwidth]{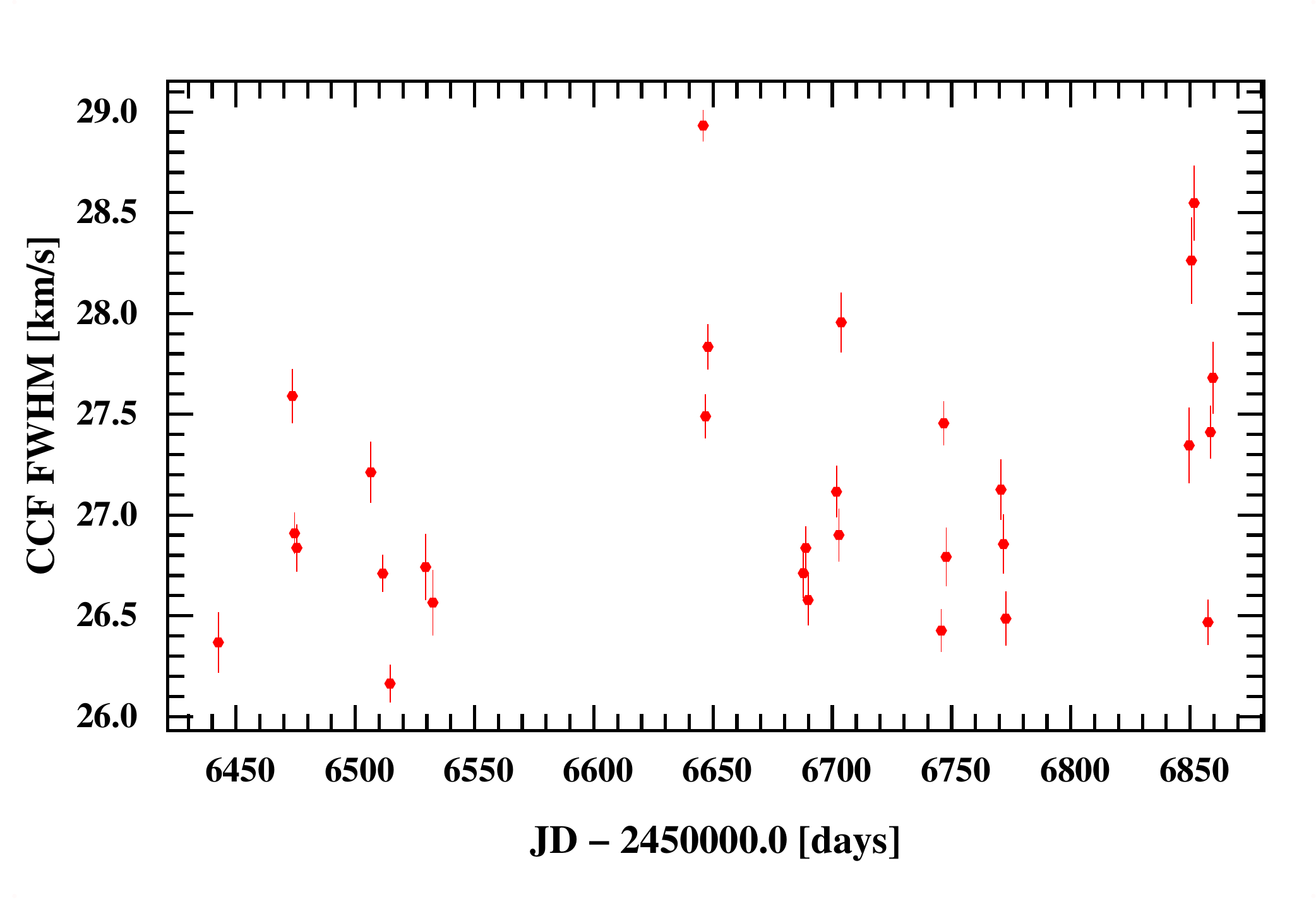}

\caption{CORALIE measurements of the FWHM of the cross
correlation function. The higher scatter in the second season is thought
to be due to an increase in activity of the star. \label{fig:cor3}
}

\end{figure}

\begin{figure}
\centering
\includegraphics[angle=0,width=\columnwidth]{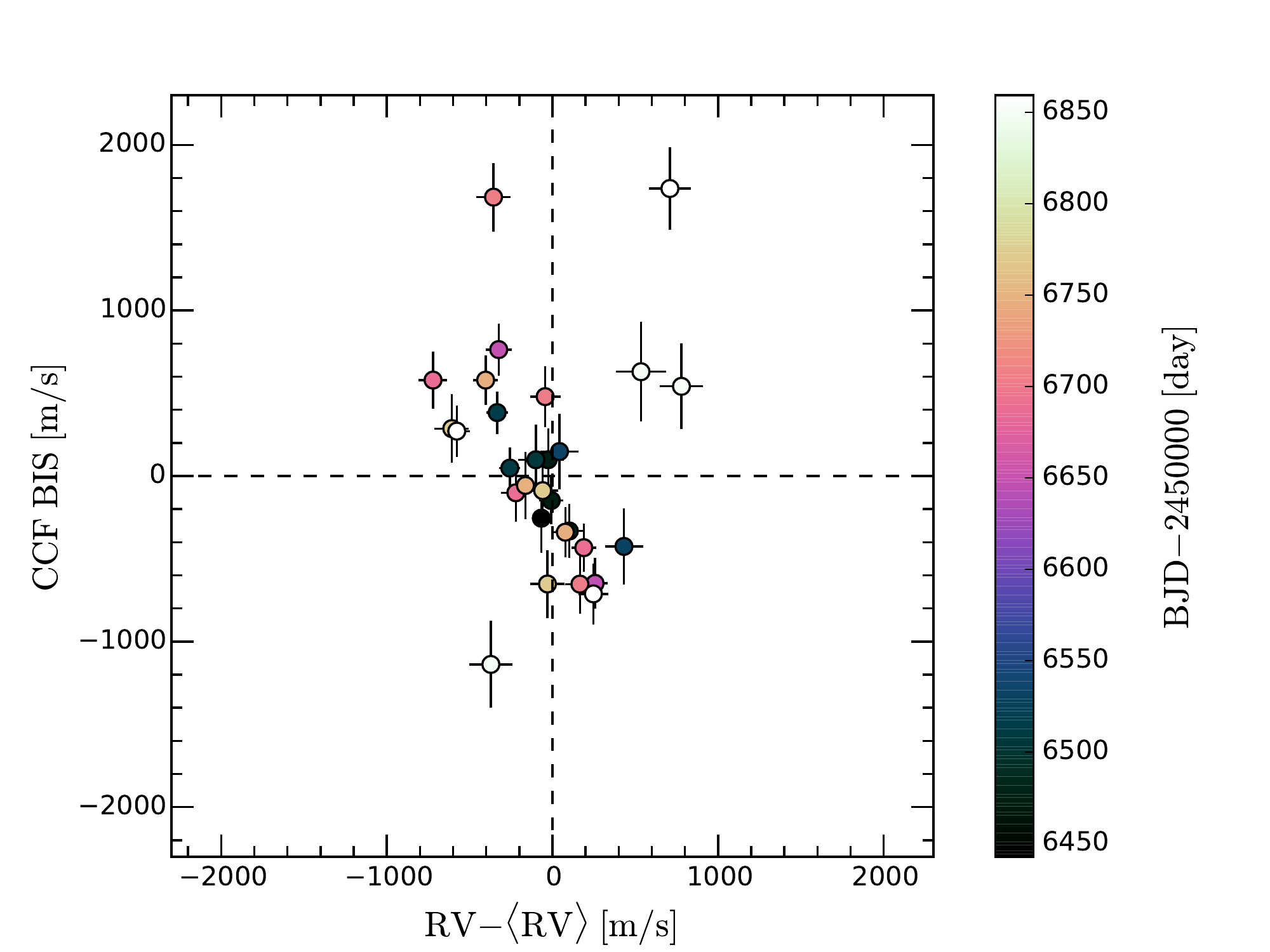}

\caption{CORALIE measurements showing the span in the slope of the
  bisector plotted against the measured RV. Both
  quantities are anti-correlated, implying that the apparent RV
  variability is caused by a change in the shape of the cross-correlation function, implying that variation in the shapes of the
  lines are due to stellar activity.\label{fig:cor4} }

\end{figure}

The first spectrum was obtained on UT 2013 May 30. It was a 3600s
exposure without simultaneous Th--Ar, to assess the observability of
the target.  We then follow the observing techniques devised in
\citet{Dumusque11} to mitigate stellar astrophysical noise:
observations were scheduled by blocks of three spectra obtained on
consecutive nights, to approximately match the rotation period of the
star \citep{Mamajek12}. 29 spectra with exposures ranging from 900 to
1800s were obtained between UT 2013 June 30 and UT 2014 July 21
(Table \ref{tab:coralie}). The observations in Fig. \ref{fig:cor1}
show significant scatter in radial velocities, but also in the slope
of the bisector (Fig. \ref{fig:cor2}) and the width of the lines
(Fig. \ref{fig:cor3}). The variation in radial velocity is strongly
anti-correlated to the slope of the bisector, a sign that stellar
activity is distorting the shape of absorption lines (see Fig.
\ref{fig:cor4}).  The scatter in the bisector slope and FWHM has
increased during the 2014 observing season compared to 2013.

\begin{table*}
\centering
 \caption{CORALIE observations of J1407\label{tab:coralie}}
\begin{tabular}{cccccc}
\hline
BJD & RV & $\sigma_{\rm RV}$ & FWHM & Bis-span & Exposure \\
(d) & (km s$^{-1}$) & (km s$^{-1}$) & (km s$^{-1}$) & (km s$^{-1}$) & (s) \\
\hline
56442.576884 & 6.888 & 0.104 & 26.368 & -0.25 & 3600\\
56473.624422 & 6.929 & 0.094 & 27.590 &  0.09 & 1800\\
56474.551068 & 6.948 & 0.071 & 26.909 & -0.14 & 1800\\
56475.489472 & 7.056 & 0.081 & 26.836 & -0.33 & 1800\\
56506.514193 & 6.854 & 0.106 & 27.212 &  0.09 & 1800\\
56511.528147 & 6.698 & 0.063 & 26.709 &  0.04 & 1800\\
56514.557647 & 6.621 & 0.064 & 26.163 &  0.38 & 1800\\
56529.495203 & 7.386 & 0.114 & 26.741 & -0.42 & 1800\\
56532.532901 & 6.997 & 0.113 & 26.565 &  0.14 & 1800\\
56645.860211 & 8.089 & 0.053 & 28.932 & -2.87 & 1800\\
56646.848954 & 7.212 & 0.076 & 27.489 & -0.64 & 1223\\
56647.848845 & 6.630 & 0.078 & 27.834 &  0.76 & 1200\\
56687.864277 & 6.234 & 0.086 & 26.711 &  0.57 & 1200\\
56688.871105 & 7.144 & 0.073 & 26.836 & -0.43 & 1200\\
56689.832639 & 6.734 & 0.088 & 26.578 & -0.10 & 1200\\
56701.740155 & 7.121 & 0.088 & 27.115 & -0.65 & 1200\\
56702.733195 & 6.911 & 0.092 & 26.900 &  0.47 & 1200\\
56703.722404 & 6.599 & 0.103 & 27.955 &  1.68 & 1200\\
56745.690237 & 6.551 & 0.074 & 26.426 &  0.57 & 1200\\
56746.714550 & 7.033 & 0.076 & 27.455 & -0.33 & 1200\\
56747.718248 & 6.792 & 0.101 & 26.792 & -0.05 & 1200\\
56770.687959 & 6.347 & 0.103 & 27.126 &  0.28 &  900\\
56771.759089 & 6.925 & 0.102 & 26.855 & -0.65 &  900\\
56772.723642 & 6.896 & 0.092 & 26.486 & -0.08 &  900\\
56849.597842 & 6.583 & 0.131 & 27.345 & -1.13 &  900\\
56850.544421 & 7.490 & 0.150 & 28.263 &  0.63 &  900\\
56851.681482 & 7.734 & 0.129 & 28.548 &  0.54 &  900\\
56857.466766 & 6.378 & 0.077 & 26.468 &  0.27 &  900\\
56858.516701 & 7.202 & 0.092 & 27.411 & -0.71 &  900\\
56859.502439 & 7.664 & 0.125 & 27.681 &  1.74 &  900\\
\hline
\end{tabular}
\end{table*}

\begin{figure}
\centering
\includegraphics[angle=0,width=\columnwidth]{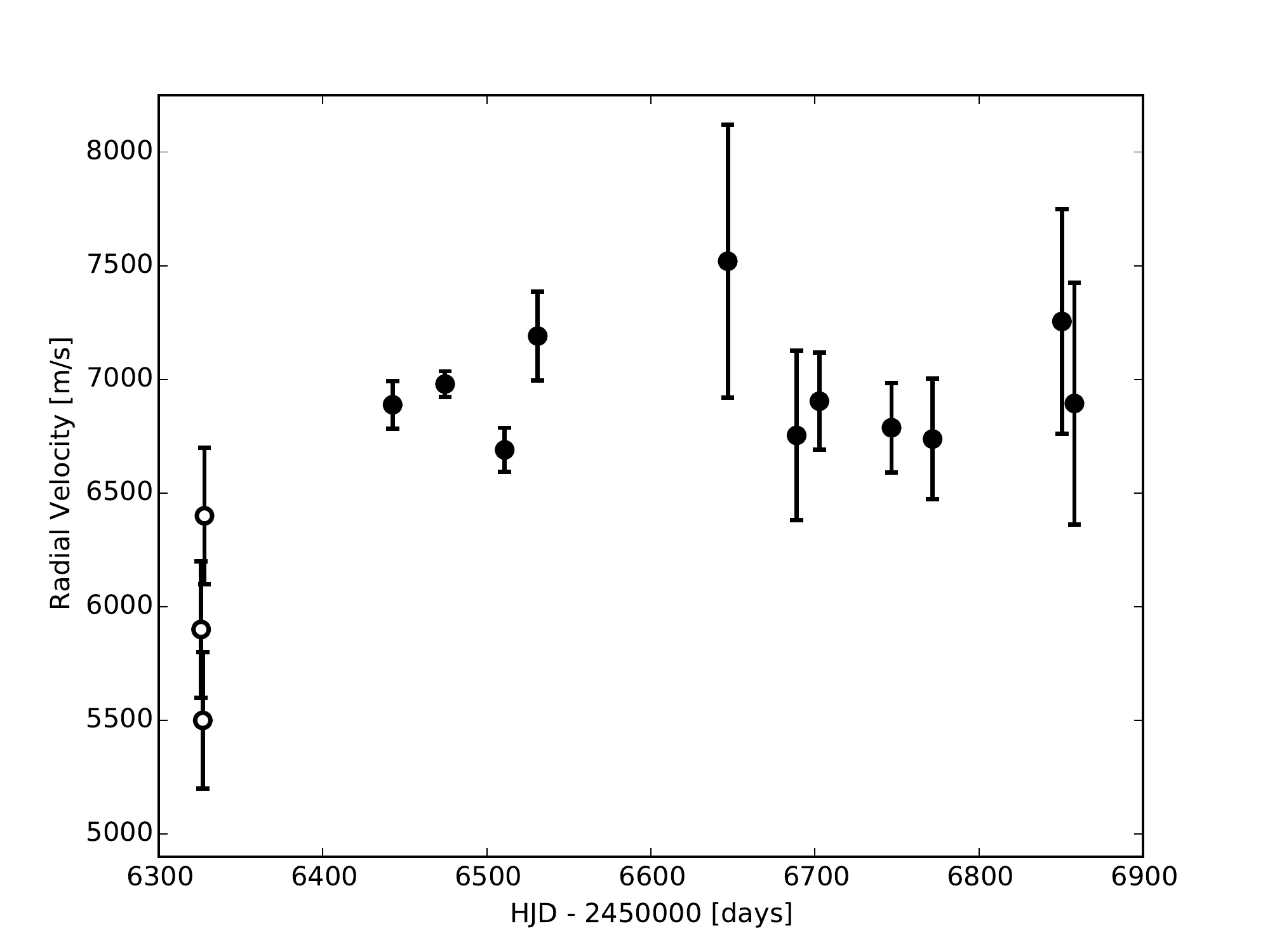}

\caption{Radial Velocity measurements of J1407. Open circles are data
  from MIKE and filled circles are binned data from CORALIE.
\label{fig:rv}}

\end{figure}

The data were combined over each observing set, reducing the sample to
nine individual measurements: the velocities were optimally averaged,
with an associated error obtained from the RMS within a set, and the
corresponding mean date calculated. The results are graphically
displayed in Fig. \ref{fig:rv}. Errors are larger in the seven epochs
taken in the 2014 observing season, reflecting an increase in the
activity of the star.

%%% PROMPT

\subsection{Photometric Observations with PROMPT\label{sec:prompt}}

\begin{figure}
\centering
\includegraphics[angle=0,width=\columnwidth]{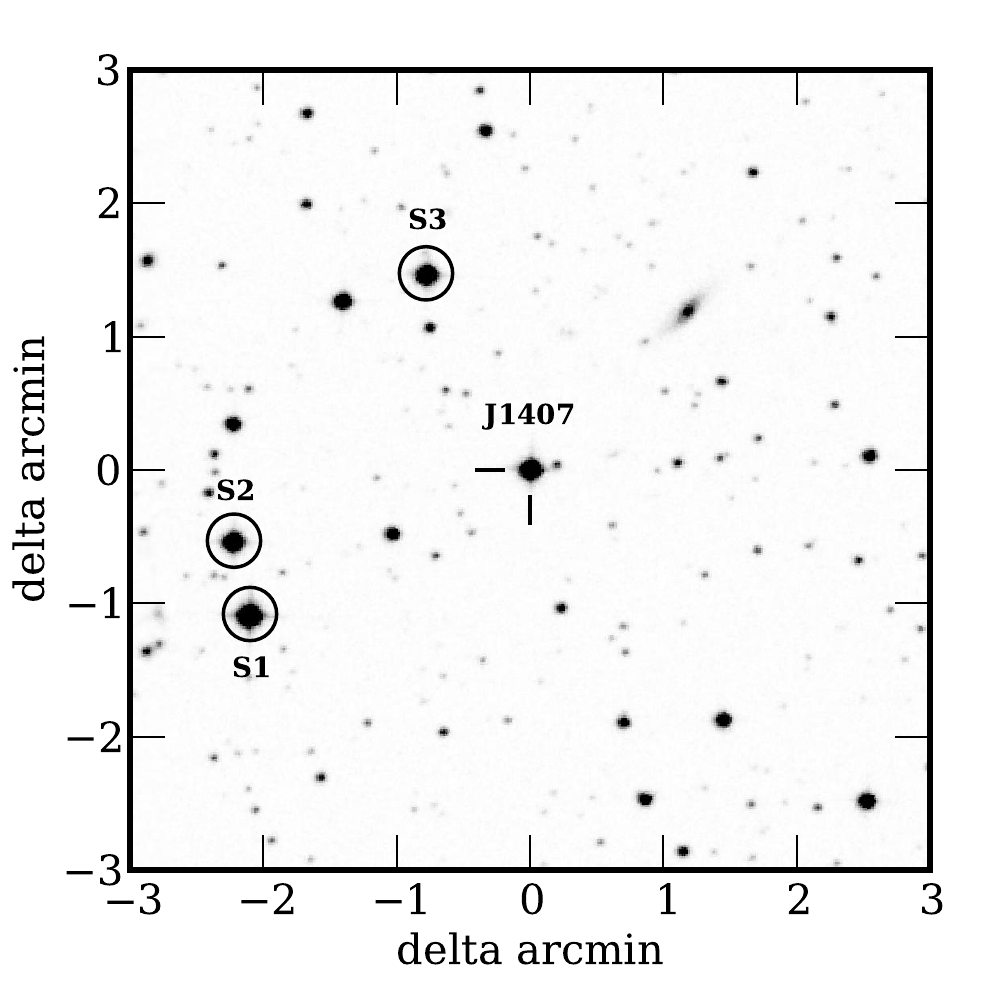}
\caption{Finder chart for J1407 and three calibration stars generated
using SkyView (2nd generation Digitized Sky Survey, red). North is up,
east is left. Field of view is 6 by 6 arcminutes.
\label{fig:finder}}
\end{figure}

\begin{figure*}
\centering
\includegraphics[angle=0,width=\textwidth]{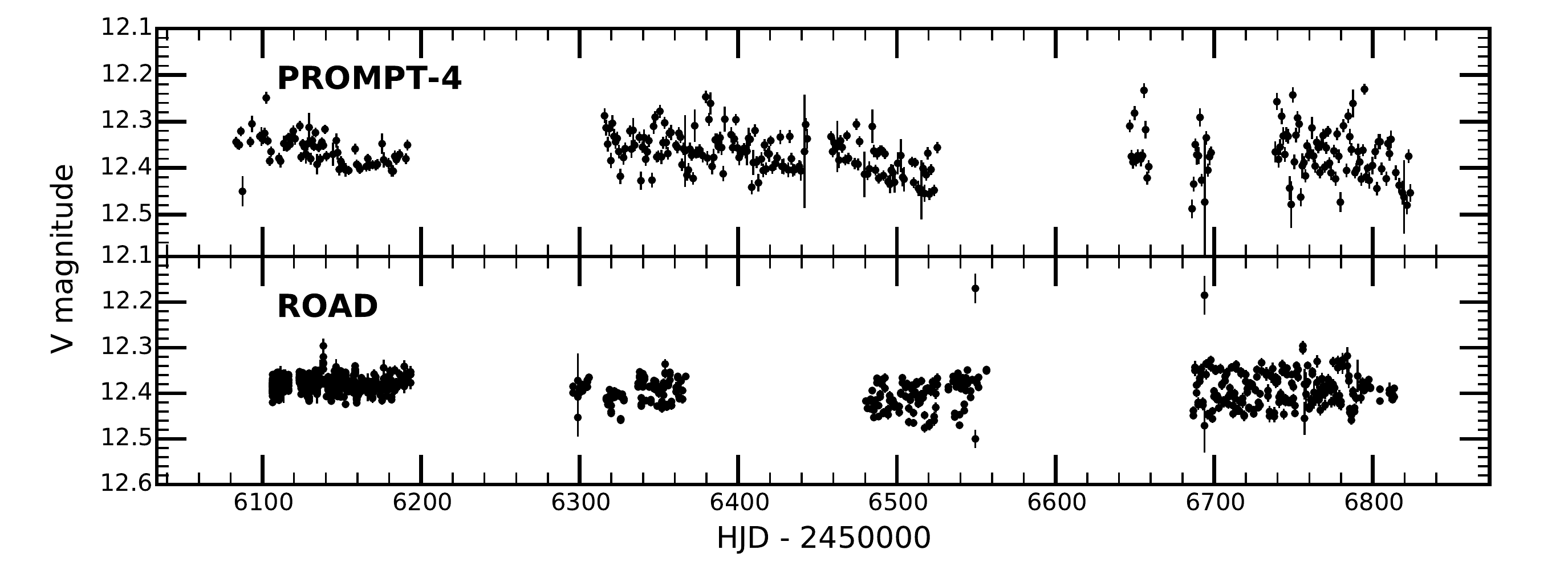}
\caption{Photometry of J1407. {\it Upper panel:} PROMPT-4 $V$-band light curve for J1407 between June 2012 and June 2014,
covering 330 nights. Photometry has been measured with respect to calibration
star S1 in Table \ref{tab:calstars}. {\it Lower panel:} ROAD photometry
of J1407 between July 2012 and June 2014.
\label{fig:PROMPT_lightcurve}}
\end{figure*}

The brightness of J1407 has been monitored nearly nightly between
2012 June and 2014 June, when possible, with the PROMPT-4 telescope at CTIO
\citep{Reichart05}.
Three exposures of 3s each are taken per clear night in both $V$ and
$I$ filters.
We report the $V$-band photometry in this contribution, which are
sufficient to determine that no long eclipses have been seen during
330 epochs during the 2012-2014 observing seasons listed in Table
\ref{tab:prompt}.
Three neighbouring comparison stars (listed in Table
\ref{tab:calstars}) have mean Johnson $V$ magnitudes observed by the
AAVSO Photometric All-Sky Survey (APASS) accurate to $\sim$0.02 mag,
and calibrated to Landolt fields \citep[Data Release 7;][]{Henden14}.
A finder chart for J1407 and the three comparison stars is provided
in Fig. \ref{fig:finder}.
From observations over 330 nights with PROMPT-4 during 2012--2014, we
find that these comparison stars are photometrically quiet to $<$0.02
mag rms, and hence provide adequate comparison stars for assessing the
variability of J1407.
Fig. \ref{fig:PROMPT_lightcurve} shows the 2012 June to 2014 June
$V$-band light curve for J1407 using calibrator star S1 for comparison.
From comparison of J1407's brightness to the three APASS comparison
stars, we find that the 2012--2014 average $V$ magnitude for J1407 was
$V$ = 12.365, with a calibration-dominated uncertainty of
$\pm$0.02 mag. 
This compares well to the median $V$ magnitudes estimated
from previous time series photometry data sets (outside of eclipse):
$V$ = 12.29 \citep[ASAS; ][580 epochs]{Pojmanski02},
$V$ = 12.34 \citep[SuperWASP; ][28194 epochs, $V_T$ converted to Johnson $V$]{Pollacco06},
and $V$ = 12.34 \citep[APASS; ][8 epochs]{Henden14}. 

A sample of the photometry is shown in Table \ref{tab:prompt_sample}.

\begin{table*}
\centering
\caption{PROMPT 2012-2014 Observing Seasons \label{tab:prompt}}
\begin{tabular}{ccc}
\hline
Start epoch & End epoch & \\
(HJD) & (HJD) & Dates \\
\hline
2456083 & 2456191 & UT 2012 Jun 04--Sep 20\\
2456315 & 2456443 & UT 2013 Jan 22--May 30\\
2456458 & 2456625 & UT 2013 Jun 14--Nov 28\\
2456646 & 2456658 & UT 2013 Dec 19--Dec 31\\
2456686 & 2456698 & UT 2014 Jan 28--Feb 09\\
2456738 & 2456823 & UT 2014 Mar 21--Jun 14\\
\hline
\end{tabular}
\end{table*}

%%% TABLE OF PROMPT PHOTOMETRY

\begin{table*}
\centering
\caption{ A sample of the photometry from the PROMPT-4 telescope. The full table is available online.  \label{tab:prompt_sample}}
\begin{tabular}{ccc}
\hline
Epoch & $V$ band & $V$ band error \\
(HJD - 2450000) & (mag) & (mag) \\
\hline
6083.46945 & 12.316 & 0.013 \\
6084.46976 & 12.316 & 0.013 \\
6085.48287 & 12.385 & 0.013 \\
6086.46948 & 12.321 & 0.013 \\
6087.46938 & 12.429 & 0.041 \\
\hline
\end{tabular}
\end{table*}

%%% FINDER CHART %%%

%%% PHOTOMETRIC COMPARISON STARS

\begin{table*}
\centering
\caption{J1407 and Photometric comparison stars \label{tab:calstars}}
\begin{tabular}{cccc}
\hline
Name & 2MASS & UCAC4 & V$_{mag}$\\ 
\hline
J1407 & J14074792-3945427 & 252-062736      & 12.336\,$\pm$\,0.034\\
S1    & J14075890-3946482 & 252-062751      & 11.238\,$\pm$\,0.016\\
S2    & J14075952-3946151 & 252-062753      & 12.150\,$\pm$\,0.017\\
S3    & J14075199-3944151 & 252-062744      & 12.390\,$\pm$\,0.024\\
\hline
\end{tabular}
\begin{minipage}{13cm}
2MASS identifiers are from \citet{Skrutskie06} and
  encode the ICRS position.  UCAC4 star identifiers from
  \citet{Zacharias12}.  Mean $V$ magnitudes and uncertainties are from
  the AAVSO Photometric All-Sky Survey (APASS) Data Release
  7. Long term monitoring of
  these stars with PROMPT-4 (330 epochs) shows that they are nearly
  constant at the $<$0.02 mag rms level.
\end{minipage}
\end{table*}

%%% ROAD

\subsection{Photometric observations with ROAD\label{sec:road}}

Nightly observations of J1407 have been taken at the  ROAD \citep{Hambsch12}
observatory in
Chile, using a commercially available CCD camera from Finger Lakes
Instrumentation\footnote{\url{http://www.flicamera.com/}}, an FLI ML16803 CCD,
 The Microline (ML) line
of CCD cameras is a lightweight camera design which can hold a variety
of CCD chips. We used a full frame Kodak KAF-16803 image sensor together
with the 40 cm Optimized Dall Kirkham $f/6.8$ telescope from Orion
Optics UK\footnote{\url{http://www.orionoptics.co.uk/}} to give a field of
view of nearly 48 arcmin $\times$ 48 arcmin, and the data were binned
$3\times 3$ for a plate scale of $2.09$ arcsec/pixel$^{-1}$ to keep the amount of data
to a reasonable value. The download speed was 8 MHz, the preamplifier
gain 1.4 e$^-$/ADU. The camera was thermoelectrically cooled to a
temperature of $-25^o$C. All the measurements are made with a photometric V
filter from Astrodon corporation. Dark and Flat frame correction is done
in the CCD camera control program
MAXIM/DL\footnote{\url{http://cyanogen.com/}}.

Data reduction is done using software developed by P. de Ponthierre
\footnote{LesvePhotometry software
  (\url{http://www.dppobservatory.net/AstroPrograms/Software4VSObservers.php})}
based on comparison stars from the AAVSO sequence for J1407.
000-BKN-848 (RA 14:08:18.78, DEC -39:49:54.1, mag 12.051) is used as a
reference star and 000-BKN-850 (RA 14:07:40.73, DEC -39:38:53.7, mag
13.102) as a comparison star. While no colour transformation was made,
the median $V$ magnitude of the observations (12.38; rms = 0.02 mag)
is within 0.02 mag of the PROMPT observations, and hence can be
considered sufficiently accurate to the other $V$-band observations
for our eclipse search. The resultant photometry is seen in Fig.
\ref{fig:PROMPT_lightcurve}.

A sample of the photometry is shown in Table \ref{tab:road_sample}.

%%% TABLE OF PROMPT PHOTOMETRY

\begin{table*}
\centering
\caption{A sample of the photometry from the ROAD observatory. The full table is available online.  \label{tab:road_sample}}
\begin{tabular}{ccc}
\hline
Epoch & V band &V band error \\
(HJD) & (mag) & (mag) \\
\hline
2456106.46175 & 12.387 & 0.008 \\
2456106.46417 & 12.397 & 0.007 \\
2456106.46658 & 12.394 & 0.007 \\
2456106.46900 & 12.406 & 0.008 \\
2456106.47142 & 12.396 & 0.007 \\
\hline
\end{tabular}
\end{table*}

\section{Revised Stellar Parameters for J1407}
\label{sec:stellar}

We update the stellar parameters for J1407 and present them in Table
\ref{tab:starparam}, together with references.

\begin{table*}
 \centering
    \caption{J1407 stellar parameters\label{tab:starparam}}
    \begin{tabular}{lll}
    \hline
Parameter & Value & Notes \\
\hline
Mass & $0.9M_\odot$ &  \citet{vanWerkhoven14} \\
Radius &  0.99\,$\pm$\,0.11 R$_{\odot}$ & This paper \\
Rotational period & $3.21\pm0.01$ day & \citet{vanWerkhoven14} \\
Minimum radius & R$_{\star} > 0.93\pm0.02$ & This paper \\
Distance & $133\pm12$ pc & \citet{vanWerkhoven14}  \\
T$_{\rm eff}$ & 4400$\pm$100\,K & This paper \\
log$(L/L_\odot)$ & $-0.478\pm0.077$ dex & This paper \\
\hline
\end{tabular}
\end{table*}

\subsection{Revised Spectral Energy Distribution} 

\begin{figure}
\centering
\includegraphics[angle=0,width=\columnwidth]{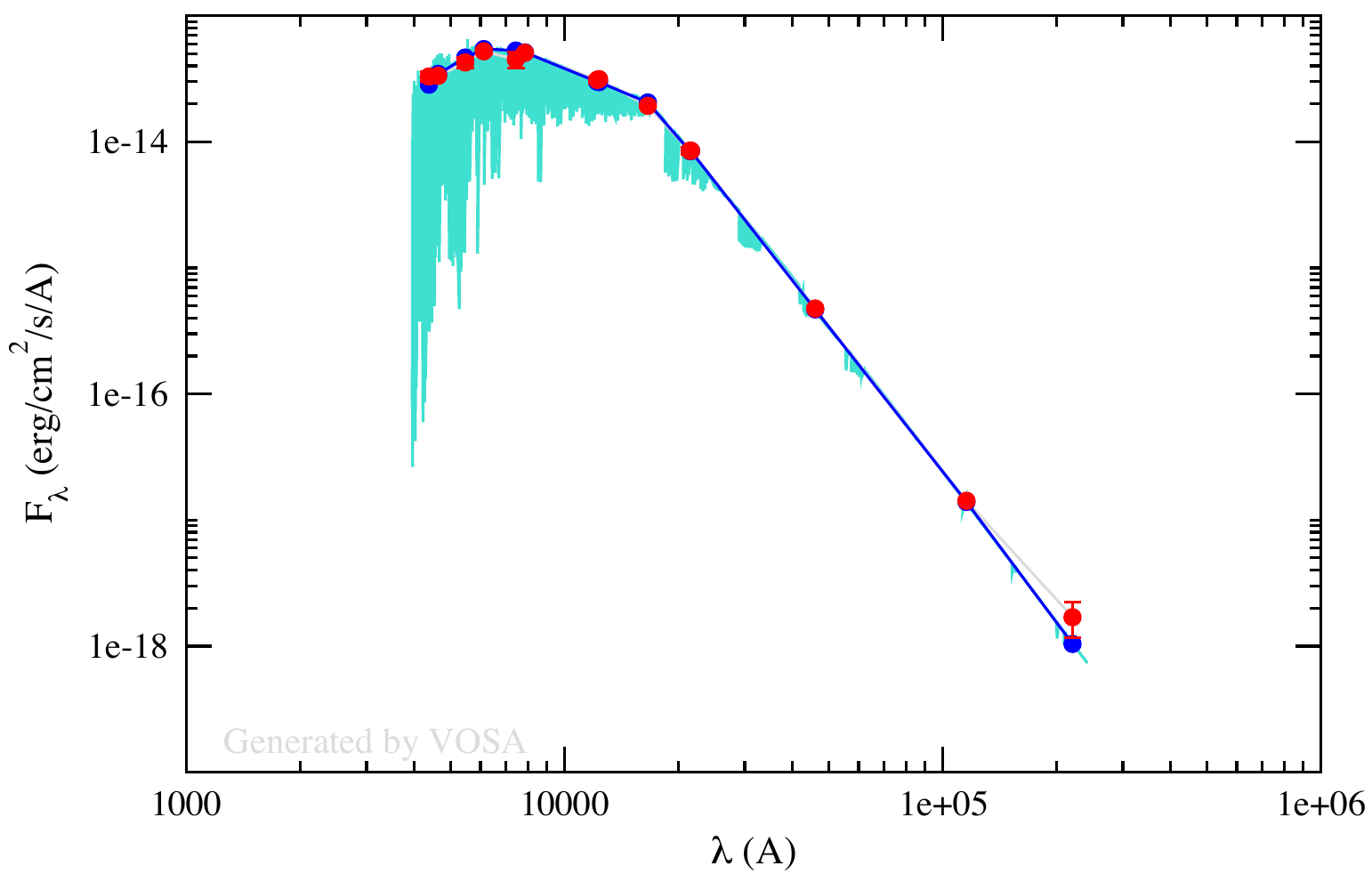}
\caption{Spectral energy distribution and best-fitting model for J1407.
Red points are the measured photometry with error bars. The dark blue
points and line are the best fit of the model to the photometric points,
and the light blue is the model spectrum.
\label{fig:SED}}

\end{figure}

A new spectral energy distribution (SED) for J1407 was constructed using the
Virtual Observatory SED Analyzer
(VOSA)\footnote{\url{http://svo2.cab.inta-csic.es/theory/vosa4/}}. Photometry
in 14 bands was used: $B$, $g'$, $r'$, and $i'$ from APASS DR7
\citep{Henden14}, $V$ from Section \ref{sec:prompt}, $I$, $J$, $K_s$ from
DENIS \citep{Epchtein97}, $J$, $H$, $K_S$ from 2MASS
\citep{Skrutskie06}, and the $W1, W2, W3, W4$ bands from All-WISE
\citep{Wright10,Cutri13}.  BT-SETTL models \citep{Allard11} with solar
metallicity over range log($g$) = 4-5 and extinction range $A_V$ =
0-0.3 were fit to the photometric data. The best fit SED (plotted in
Fig. \ref{fig:SED}) has T$_{\rm eff}$ = 4400\,K, log($g$) = 4.0,
[$\alpha/Fe$] = 0.2, negligible extinction ($A_V$ = 0.00). The T$_{\rm
  eff}$ inferred from the SED fit is consistent with previous
estimates \citep{Mamajek12,vanWerkhoven14} and the spectral type
estimate (K4.8) from the MIKE analysis (\S2.4). The SED bolometric
flux is (5.974\,$\pm$\,0.020) $\times$ 10$^{-11}$ erg\,s\,cm$^{-2}$,
or on the IAU scale an apparent bolometric magnitude of
11.577\,$\pm$\,0.037 mag. At the revised kinematic distance of
133\,$\pm$\,12 pc \citep{vanWerkhoven14}, the revised luminosity is
log(L/L$_{\odot}$) = -0.478\,$\pm$\,0.077 dex (with the uncertainty
dominated by the distance uncertainty).

\subsection{Stellar Equatorial Velocity, Stellar Radius, and Inclination}

We can place some further constraints on the size and inclination of
the star J1407.  Adopting T$_{\rm eff}$ = 4400$\pm$100\,K from the SED
fitting (consistent with previous estimates), the updated luminosity
converts to a new stellar radius estimate of 0.99\,$\pm$\,0.11
R$_{\odot}$. With the revised T$_{\rm eff}$ and radius, and a
period of 3.21\,$\pm$\,0.01 d \citep{vanWerkhoven14}, we predict the
star's equatorial rotation velocity to be $v_{eq}$ = 15.7\,$\pm$\,1.7
km\,s$^{-1}$.

A lower limit to the stellar radius of J1407 can be estimated directly
from the projected rotational velocity (\S2.4) and the stellar
rotation period. Taking $v$\,sin\,$i_{\star}$ = 14.6\,$\pm$\,0.4 km\,s$^{-1}$
and $P$ = 3.21\,$\pm$\,0.01 d, we estimate a strong lower limit to
the stellar radius of R$_{\star}$ $>$ 0.93\,$\pm$\,0.02 R$_{\odot}$.

Since we have estimates of the equatorial and projected rotational
velocities, we can estimate the inclination of the star J1407.  We ran
a Monte Carlo simulation taking into account the stellar parameters
and their uncertainties to estimate the probability distribution of
the stellar inclination $i_\star$. For 10$^6$ simulations, 73.6\%\, resulted
in physical solutions ($i_\star$ $<$ 90$^{\circ}$). We estimate the stellar
inclination to be $i_{\star}$ = ${63.9^{\circ}}^{+10.5^{\circ}}_{-8.5^{\circ}}$
  (68.3\%\, CL) or $^{+19.9^{\circ}}_{-15.0^{\circ}}$ (95.5\%\,CL).

\section{Analysis}
\label{an}

We determine the constraints that these null detections provide by
modelling orbits for J1407b as a function of orbital period.
We assume that J1407b is gravitationally bound to J1407 and follows
Keplerian orbits parameterized by orbital elements $a$, $e$, $i$, $P$,
$T$, $\omega$, $\Omega$.
The centre of the J1407b transit was UT 2007 April 27, and in the absence of any other perturbing bodies in the
system, we assume that it follows a closed elliptical orbit with
inclination $i=90^o$ and ellipticity $e$.

We determine the mass and period limits for the case of circular orbits,
and for the more generalized case of elliptical orbits in Section
\ref{sec:ell}.
For the total mass function of the system $M=0.9 M_\odot$ we investigate
trial periods $P$ from 2 to 1000 yr.

\subsection{Circular orbits}

\subsubsection{Limits from ring geometry and orbital velocity for circular orbits}

\begin{figure*}
\centering
\includegraphics[angle=0,width=\textwidth]{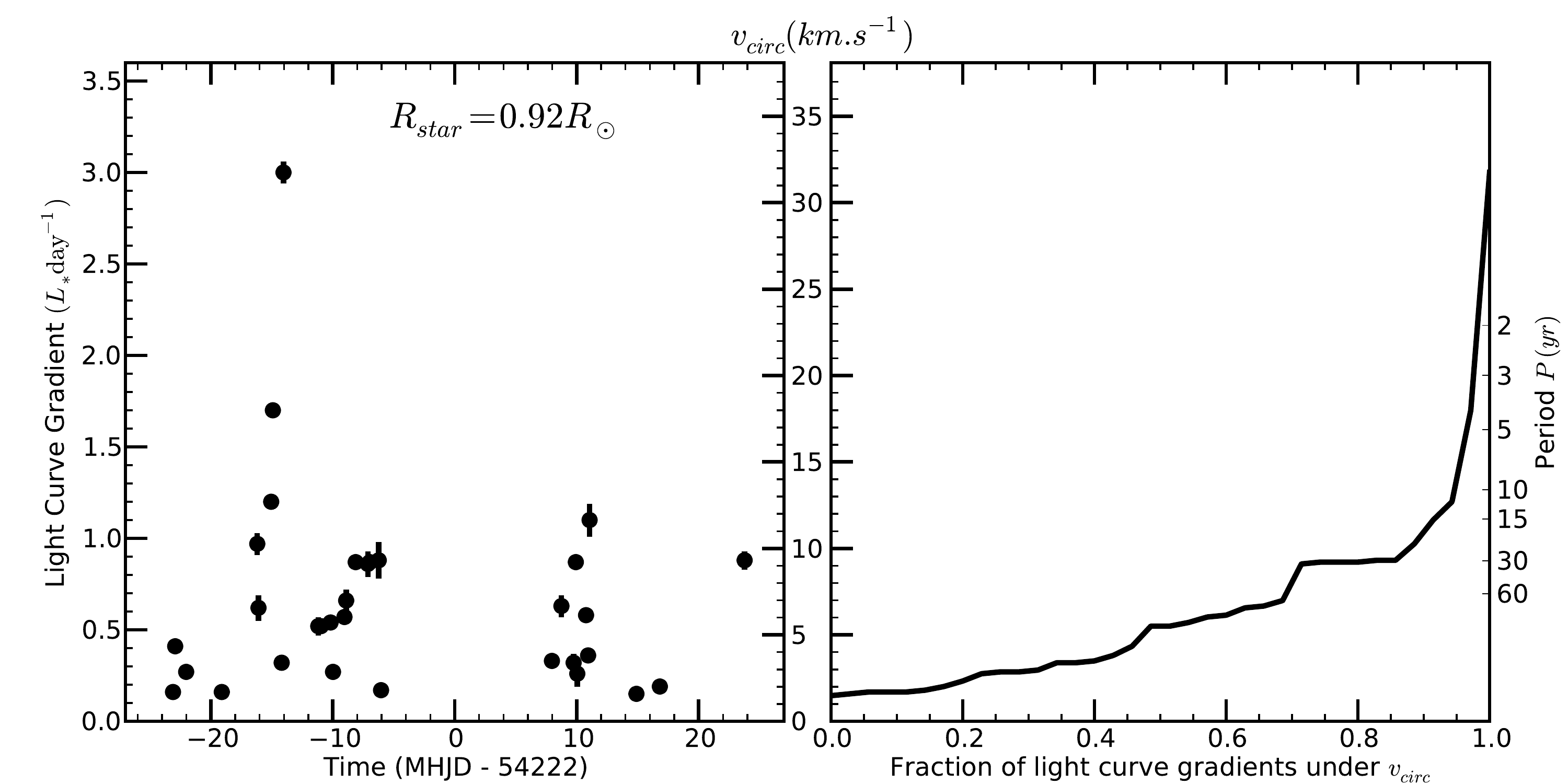}

\caption{Gradients of the J1407 light curve as a function of time. The
gradients are taken from \citet{vanWerkhoven14} and a stellar radius
$R_*=0.92R_\odot$ adopted. The right hand scale shows the transverse
velocity required to produce that gradient. \label{fig:gradients}}

\end{figure*}

The companion is assumed to orbit J1407 with a circular orbital
velocity $v_{circ}$ which can bring a ring edge across the disk of the
star, causing it to dim at a rate determined by the diameter of the
star.
In \citet{vanWerkhoven14}, the light curve of J1407 is analysed and the
minimum circular velocity required to match the observed gradients is
determined.
We recalculate the orbital velocities required for our determined
minimum stellar radius from Section \ref{sec:stellar} and plot the
results in Fig. \ref{fig:gradients} in the left hand panel.

For increasing orbital period $P$, the circular orbital velocity becomes
slower, and it is impossible to bring an opaque occulter across the disk
of the star fast enough to explain the measured light curve gradients
seen in the J1407 data.
The right hand panel shows the fraction of measured gradients that can
be explained by a dark or grey occulter moving a given circular orbital
velocity, along with the corresponding orbital periods on the right hand
axis.
It is clear, however, that decadal orbital periods are ruled out since
they fail to explain many of the measured gradients. 
We consider that long-period solutions where the companion has orbital
velocity $v_{orb}$ $<$ 12 km\,s$^{-1}$ ($P$ $\gtrsim$ 13.8 yr; $a$
$\gtrsim$ 5.5 au) are strongly rejected by the presence of the steep
light curve gradients discussed in \citet{vanWerkhoven14}.

\subsubsection{Limits from Direct Imaging}
\label{sec:dir}

\begin{figure*}
\centering
\includegraphics[angle=0,width=14cm]{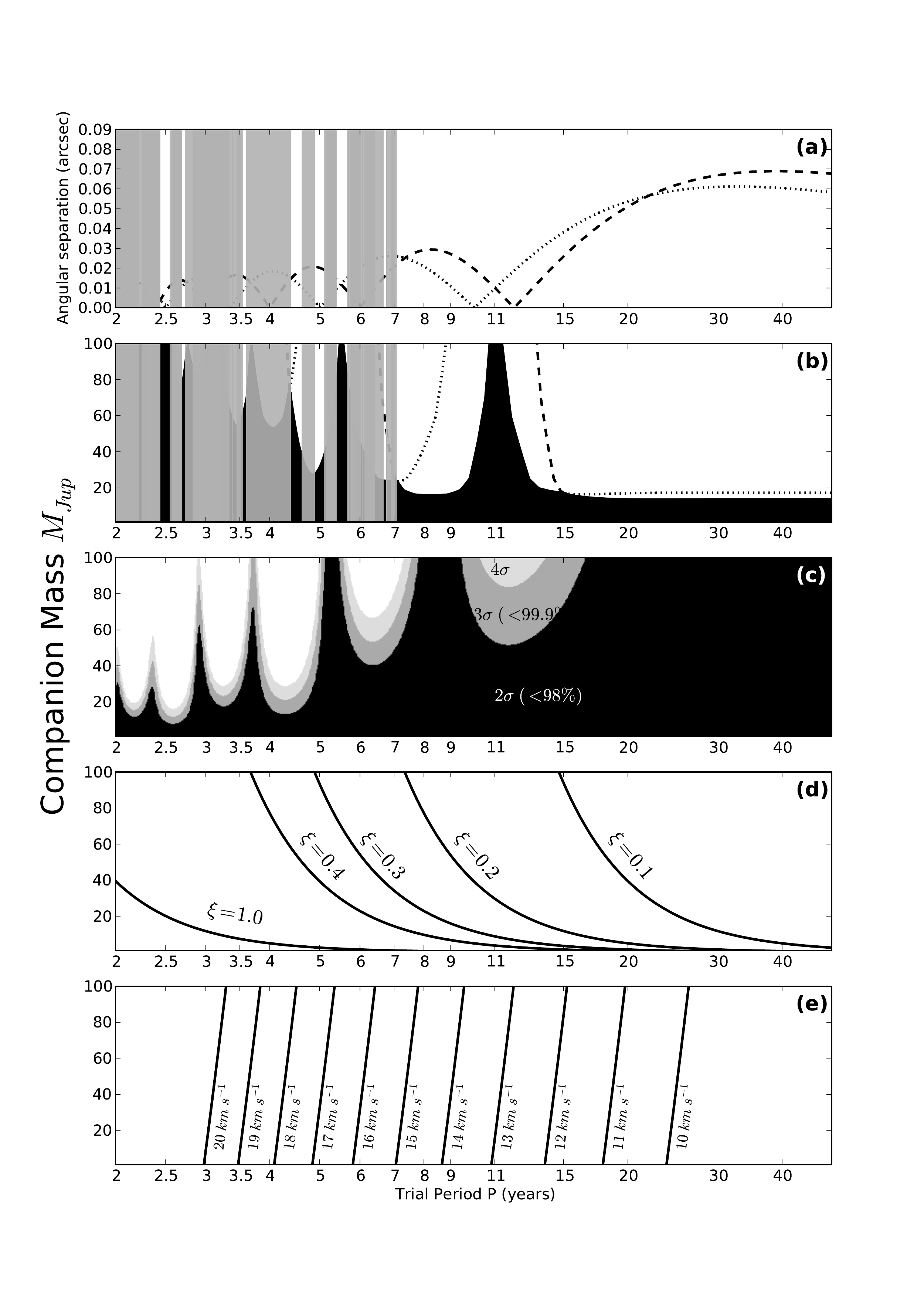}

\caption{Limits to the period and mass of J1407b for circular orbits. Panel (a) shows
  projected angular separation as a function of orbital
  period $P$ for J1407b for the Keck and VLT SAM imaging data assuming
  circular edge-on orbits with $r=0$ at UT 2007 April 30.
(b) Angular separation converted to upper mass limit.
(c) Values of goodness of fit of the RV model expressed in standard
deviations - black, dark grey and light grey are $2,3,4\sigma$
respectively.
Vertical grey bars are trial periods that have been ruled out by
photometric monitoring.
(d) Mass limits from Hill radius of the secondary companion
and the duration of the eclipse. Panel (e) shows the circular orbital
velocity $v_{circ}$ of the companion relative to the primary.
\label{periodsep}}

\end{figure*}

We calculate the projected angular separation in arcseconds as a
function of orbital period for the epochs of our two direct imaging
observations UT 27 March 2013 and UT 04 April 2012, and these are
plotted in Fig. \ref{periodsep} in Panel (a).
For an observation at a specific epoch, the projected angular
separation varies as a function of trial period. The largest angular
separation occurs when a putative companion reaches
quadrature, and is zero when $\Delta t = nP/2$, where $P$ is the
period, $n$ is a positive integer and $\Delta t$ is time between the
epoch of observation and the midpoint of the eclipse, which we take to
be UT 29 April 2007.

We do not detect any point sources in either epoch. The resultant
sensitivity maps are in delta magnitudes, which when combined with the
distance and age of the stellar system, are converted using BT-SETTL
models \citep{Allard12} into upper mass limits as a function of position
on the sky. Azimuthally averaging these limits results in a contrast
curve of upper companion mass versus distance from the J1407 in
astronomical units (see Figs \ref{samsense} and \ref{fig:kecksam}).
Projected angular separation is then converted into companion upper mass
limit in Fig. \ref{periodsep} (b). The upper mass limits are
indicated by a dotted line for the Keck image and a dashed line for the
VLT image. The black region represents the upper mass limit from either
Keck or VLT, whichever one gives the lower mass sensitivity at a given
period.

\subsubsection{Limits from photometric monitoring}

J1407 underwent a series of nearly time-symmetric eclipse events with
the midpoint of the eclipses on HJD 2454220 during early 2007. No
other eclipses of J1407 are observed over the nine year span of data
from WASP-South and ASAS as reported in \citet{Mamajek12}, with a period
search excluding all periods $P<850$ d. The incompleteness of the
photometry of J1407 is discussed in \citet{Mamajek12} and presents
orbital periods that are not ruled out by the photometric data.

In the new 2012-2014 time series photometry data from PROMPT-4 and ROAD, we see
no new deep eclipses analogous to that seen in 2007.
However, the recent PROMPT and ROAD photometric monitoring has helped rule out
a wide range of periods of duration $\sim$4 to $\sim$7 years.
The most recent photometry is now excluding orbital periods that are a
half and a third of the latest and most recent baselines, notably the
window at 3.5 years.
Periods that are ruled out are
represented by the dark grey vertical bars on the left hand side of
Fig. \ref{periodsep} in panels (a) and (b).

\subsubsection{Limits from RV Measurements}
\label{rvdir}

The RV measurements are shown in Fig.
\ref{fig:rv}.  These measurements are a combination of observations
from both MIKE and CORALIE. The star is a rapid rotator as determined
by photometric variability with a periodicity of 3.2 d
\citep{vanWerkhoven14}. 
No known high-precision RV measurements of J1407 were
taken at the time of the eclipse in 2007.  We therefore construct a RV
model $f(P,M)$ for J1407b using circular edge-on orbits, assuming a trial
period $P$ and secondary companion mass $M$ to calculate the expected RVs at the epochs of
observation $v_{radial} = f(P,M) + C$, where $C$ is constant offset that accounts
for the unknown RV at the time of the 2007 transit. This
model is then fit to the observed RVs with $C$ as a free parameter
that minimises the $\chi^2$ of the resultant model fit.

The MIKE observations have a time baseline of less than one week, and
on their own do not produce significant constraints to our model. The
combination of RV data from different telescopes with different
instruments and RV data reduction pipelines introduces
unknown systematic errors that are difficult to quantify. The CORALIE
data covers a baseline of over one year and typically has precision
better than, or similar to, the MIKE data. The velocity zero-points
should agree at the $<$200 m\,s$^{-1}$ level, given the observed
offsets seen between CORALIE velocities and those from the California
Planet Search \citep[those used in the MIKE analysis;
  e.g. ][]{Chubak12}. A larger effect is the velocity jitter
introduced due to plage and starspots, which could easily generate
velocity variations at the tenths of km\,s$^{-1}$ for a star rotating
as fast as J1407 \citep[e.g.][]{Jeffers14}.  By restricting our
analysis to the CORALIE data, we do not need to consider the absolute
calibration of the RV (expressed in the constant $C$ in
our model) nor the systematic offsets between the two instruments.

No significant RV variation is detected for the star. A
fit to a horizontal line produces a reduced $\chi^2_{\rm r} = 1.1 \pm
0.5$, and mean systemic velocity $\gamma = 6.91 \pm 0.04$ km
s$^{-1}$. A slope produced a worse $\chi^2_{\rm r}$. At 95\%
confidence, $|\,\dot{\gamma}\,| < 450 $ m s$^{-1}$ yr$^{-1}$. The
CORALIE data alone excludes additional companions with masses $>$ 12
\,\mjup\, on circular orbits shorter than one year (but longer than 3
d). The grid of $\chi^2_{red}$ converted to goodness of fit in
$\sigma$ is shown in Fig. \ref{periodsep} (c).

Using the Upper Cen--Lup (UCL) subgroup velocity vector from
\citet{Chen11} ($U, V, W$ = -5.1\,$\pm$\,0.6, -19.7\,$\pm$\,0.4,
-4.6\,$\pm$\,0.3 km\,s$^{-1}$), we predict the RV of an
ideal UCL member at the position of J1407 to be $\gamma$ = 7.0
km\,s$^{-1}$, with a predicted rms scatter amongst members of $\pm$1.3
km\,s$^{-1}$. The measured RV is consistent with
J1407's membership in the UCL subgroup of
Sco--Cen, as proposed by \citet{Mamajek12}. Hence, the position, proper
motion, excess lithium, strong X-ray emission, HR diagram position,
and now RV, for J1407 are all consistent with UCL
membership.

\subsubsection{Limits from Dynamical Constraints of the Ring System}

The duration of the eclipse event in 2007 sets a dynamical constraint
based on the stability of the ring system around J1407b. The longest
transit time possible is one where the star J1407 passes through the
middle of the disk and behind the secondary companion. The
measured transit duration $t_{d}$ therefore represents a lower limit
on the total projected diameter of the ring system for nonzero impact
parameters of primary star.  Following the terminology of
\citet{Mamajek12}, we define the radius of the disk system as measured
from J1407b as $r_{d}$. The radius where the gravity of J1407b is
dominant over the primary star is defined as the Hill radius $r_{{\rm H}}$.
Defining the ratio $\xi = r_{d}/r_{{\rm H}}$ so that we can express the disk
radius in terms of Hill radius, we can then calculate the expected
$\xi$ for a given orbital period and mass of J1407b.

The radius of the disk is related to the eclipse duration by:

\[
 r_{disk} = \pi a t_{d} P^{-1}
\]

Taking $r_H = a \left ( \frac{M}{3M_*} \right ) ^\frac{1}{3}$ and the mass
of the companion J1407b as $M$ we then have:

\[
M = 3M_* \left ( \frac{\pi t_{d}}{\xi P } \right )^3
\]

$M_*$ is the mass of the star J1407 `A', $a$ is adopted semimajor
axis (although we are assuming circular orbits), and $P$ is the
orbital period.  Lines of $\xi$ for values of 0.1, 0.2, 0.3, 0.4 and
1.0 are plotted in Fig. \ref{periodsep} (d). A value
of 1.0 represents a disk system completely filling the Hill sphere and
presents an extreme upper limit on the potential minimum mass required
for J1407b. \citet{Martin11} suggest a tidal truncation limit of $\xi
\sim 0.4$ and hydrodynamic simulations of planets in circumstellar
disks find $\xi \sim 0.3$ \citep{Quillen98,Ayliffe09}. For the
Galilean satellites forming in the Jovian circumplanetary disk after
dissipation, smaller values of $\xi \sim 0.1-0.2$ are found
\citep{Canup02,Magni04,Ward10}. Values of $\xi$ may therefore range
from 0.1 up to 1.0, with a more precise value determined by subsequent
detection of the next eclipse. However, based on theoretical
predictions, we consider $\xi$ $>$ 0.4 solutions exceedingly unlikely.

\subsubsection{Combined Mass Period limits for J1407b for circular orbits}

\begin{figure*}
\centering
\includegraphics[angle=0,width=\textwidth]{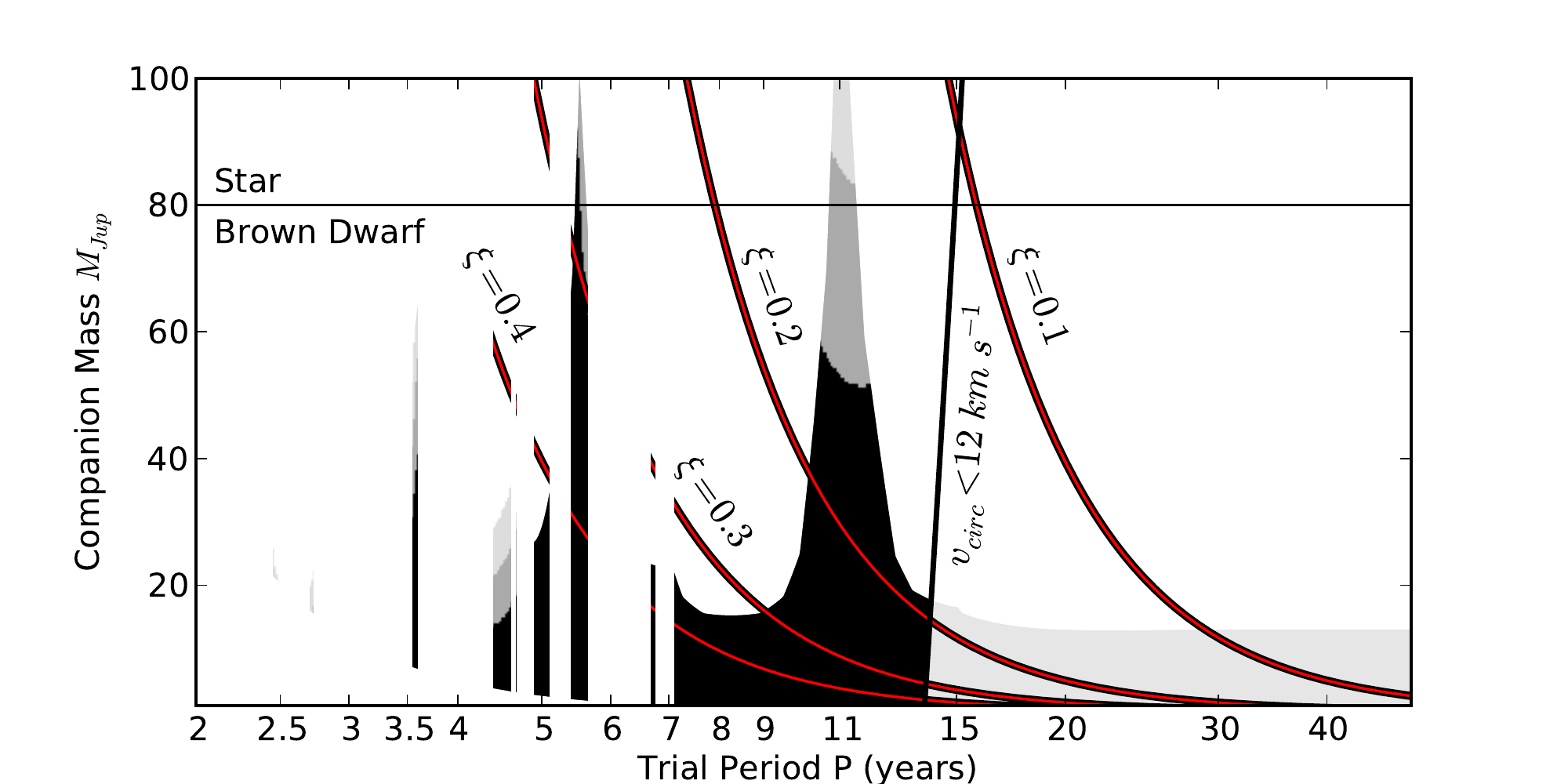}

\caption{Combined mass/period limits for circular orbits of J1407b. This figure is a
  logical combination of the panels (b), (c) and (d) in Figure
  \ref{periodsep}.
\label{fig:pemassfinal}}

\end{figure*}

Combining the four sets of constraints listed previously, we show our
limits on the possible mass and orbital period for circular orbits of J1407b in Figure
\ref{fig:pemassfinal}. Possible periods and masses are indicated by
shaded regions in the figure. Overplotted are different
values for $\xi$, the size of the ring system in units of the Hill
sphere.  Values of $\xi > 1$ are unstable and are ruled
out. Simulations indicate that $\xi > 0.4$ are unlikely and that
typical values are expected to be in the range of $0.1 < \xi < 0.3$
\citep{Quillen98,Canup02,Magni04,Ayliffe09,Ward10}.

The figure shows that the mass of the secondary is substellar
\citep[below 80\,\mjup\, for solar composition;
][]{Saumon08}\footnote{\citet{Saumon08} estimate the H-burning limit
  (defined by stars that stably fuse H at age 10 Gyr) as 0.075\,\msun,
  or $\sim$78.6 \mjup. We simply adopt the rounded value 80\,\mjup\,
  as a conservative estimate of the H-burning limit.} for many
periods, with the exception of a region of stellar mass at 5.5 yr
and 11 yr. The radial velocity measurements strongly constrain
masses at 11 years, but only slightly constrain at the 5.5 yr
period.
We can estimate the probability that the companion is above the
deuterium burning limit by using the $\chi^2$ fit from the radial
velocity model and the constraints from the non-detections of the
measurements presented in this paper.
We evaluate the $\chi^2$ of the RV model fit for a linear sampling in
companion mass from 1--100\,\mjup, and a logarithmic
sampling in period from 2--15 yr (corresponding to
$v_{circ} \approx 12$km.s$^{-1}$). The
probability $Prob$ at each sample point of mass and period is calculated using
$Prob(M,P)\,\propto\,\exp(-\chi^2/2)$, and then we normalize $Prob(M,P)$
by summing over all masses and periods and setting this sum to unity.

By integrating this grid of probability over ranges of masses, we can
determine how likely it will be that J1407b lies within a given mass
range.
Taking stellar ($M>80\mj$), brown dwarf ($80\mj > M > 13\mj$) and
planetary ($M<13\mj$) masses and integrating over all periods, we
calculate the normalized probabilities as listed in Table
\ref{tab:prob}. We explore the probabilities for different values
of the disk size in terms of Hill radius $\xi$.
The probability that J1407b is a stellar mass object is less than
$0.1\%$ for all values of $\xi$, corresponding to a $3.5\sigma$
likelihood. We can conclude that J1407b is
therefore a substellar object. For all values of $\xi$ greater than 0.3,
J1407b is equally likely to be a brown dwarf or a planetary mass object.

\begin{table*}
 \centering
 \begin{minipage}{140mm}
    \caption{Probable mass of J1407b with circular orbits\label{tab:prob}}
    \begin{tabular}{lccccc}
    \hline
Mass range                & $\xi<1.0$ & $\xi<0.5$ & $\xi<0.4$ & $\xi<0.3$ & $\xi<0.2$\\
\hline
`star'        $P(M>80\mj)$  & 0.04\% & 0.05\% & 0.07\% & 0.15\% & 0.02\%\\
`brown dwarf' $P(13-80\mj)$ & 40\%   & 48\%   & 53\%   & 59\%  & 99\%\\
`planet'      $P(<13\mj)$   & 60\%   & 52\%   & 47\%   & 41\%  & 2\%\\
\hline
Probable period $\overline{P}$ (yr)      &  9.0   &  9.5   & 10.2  & 12.1  & 12.6 \\
Probable mass   $\overline{M}$ (\mjup)   & 13.6   & 15.6   & 17.1  & 16.1  & 26.1 \\
\hline
\end{tabular}
\end{minipage}
\end{table*}

\subsection{Elliptical Orbits}
\label{sec:ell}

The light curve of J1407 shows rapid variation during the 2007 eclipse,
and regardless of the detailed large scale structure in orbit around
J1407b, the largest gradients in the light curve imply a transverse
velocity of 32 km.s$^{-1}$ for the occulting material.
If the occulting material is in orbit around J1407b itself and is
not azimuthally symmetric about J1407b then it is possible that the velocity of
this clumped material vectorially adds with the orbital velocity of J1407b to
produce the resultant transverse velocity that we see in the light curve
-- see \citet[][ Section 6.2]{vanWerkhoven14} for a discussion.
Here we explore the limiting case for elliptical orbit solutions to
J1407b where the transverse velocity is entirely due to the orbital
velocity of J1407b.
Together with the circular case in the previous section, these two
models then bracket the whole possible range of orbital solutions for
J1407b.

\subsubsection{Limits from ring geometry and orbital velocity for elliptical orbits}

As seen in the previous section, there are no possible circular orbits that have
a high enough orbital velocity to explain the highest light curve gradients.
Instead, we will assume that all transverse velocity is due to orbital
motion of J1407b at periastron, setting $\omega$ and $\Omega$
appropriately for periastron at primary transit.

The companion itself is assumed to cross in front of J1407 with a
transverse velocity $v$ which can bring a ring edge across the disk of
the star, causing it to dim at a rate determined by the diameter of the
star.
In \citet{vanWerkhoven14}, the light curve of J1407 is analysed and the
minimum transverse velocity required to match the observed gradients is
determined.
We recalculate the transverse velocities required for our determined
minimum stellar radius from Section \ref{sec:stellar} and plot the
results in Fig. \ref{fig:gradients} in the left hand panel.
The largest transverse velocity required is $32\pm2$ km.s$^{-1}$.
The largest orbital
velocity $v_{peri}$ is at periastron:

$$ v_{peri} = \frac{2\pi a}{P}\left ( \frac{1+e}{1-e} \right )^{1/2} $$

\begin{figure*}
\centering
\includegraphics[angle=0,width=15cm]{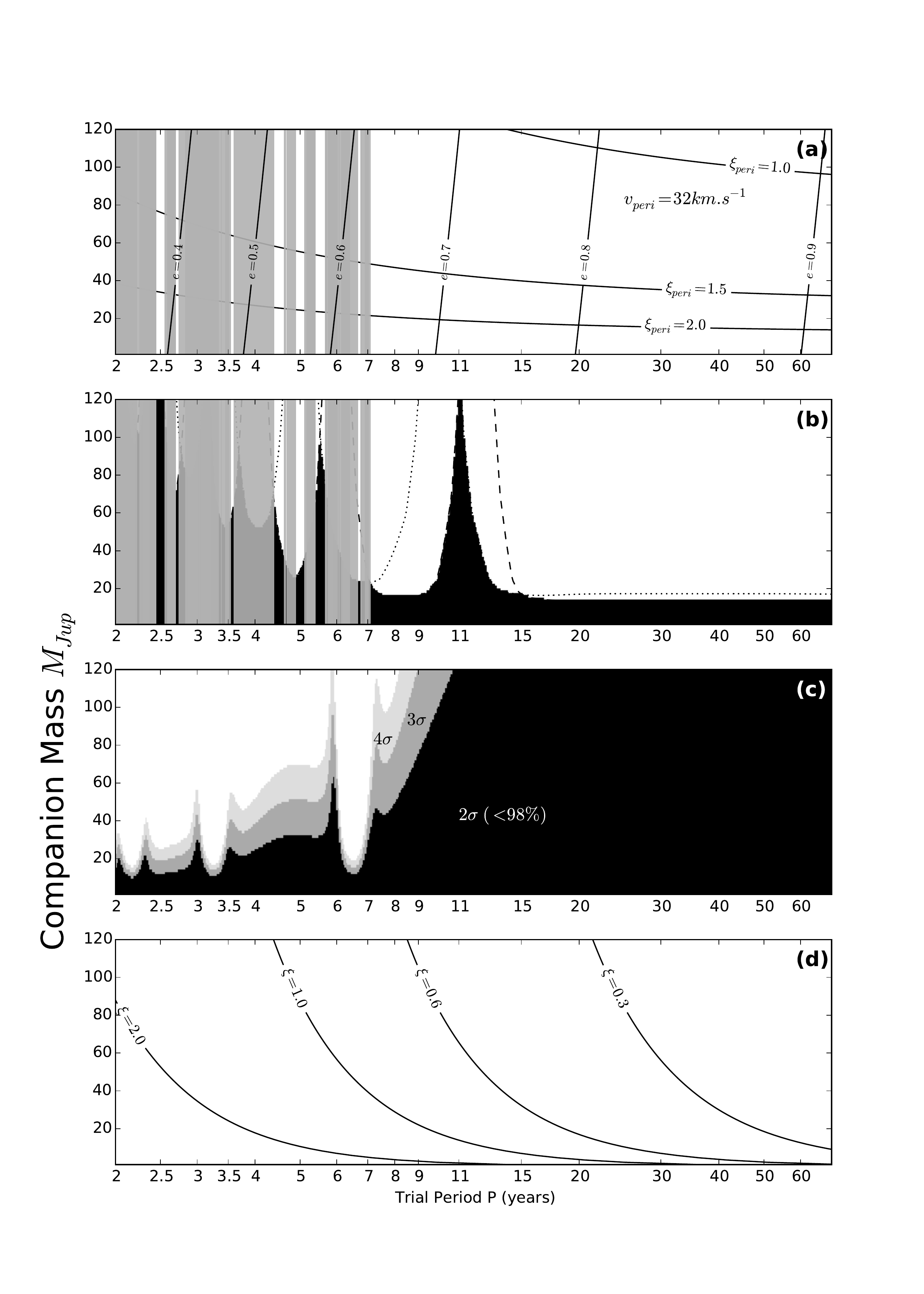}

\caption{Limits to the period and mass of J1407b for elliptical orbits
with periastron at the time of the eclipse in 2007.
Vertical grey bars are trial periods that have been ruled out by
photometric monitoring.
(a) The minimum eccentricity of orbit required to get a periastron
tangential velocity of $32km.s^{-1}$. Diameter of the ring system in
units of the Hill sphere at periastron are labelled with $\xi_{peri}$.
(b) Direct imaging limits from
the two epochs of direct imaging. The black region indicates
allowable values of mass $M$ and period $P$.
(c) Values of goodness of fit of the RV model expressed in standard
deviations - black, dark grey and light grey are $2\sigma,3\sigma,4\sigma$
respectively.
(d) The diameter of the ring system in units of the Hill sphere
calculated for the mean star/companion separation $a(1+e^2/2)$.
\label{fig:ellipt}}

\end{figure*}

If we assume that periastron occurs during the transit of J1407b, we can
then determine the minimum eccentricity required to attain a transverse
velocity of $v_{peri}$.
For values of $a$ and $P$ with $v_{peri}$ to $32\pm2$ km.s$^{-1}$, the
resultant lines of minimum eccentricities is seen in Fig. \ref{fig:ellipt} (a).

\subsubsection{Limits from Direct Imaging for elliptical orbits}

Using values of $M$, $P$ and $e$ we calculate the expected
angular separation of J1407b at the two epochs of direct imaging
observations.
The angular separation at the two epochs for each mass and period is then converted into a
direct imaging minimum detectable mass. If this mass is greater than the
companion mass $M$, then that specific mass $M$ would have been detected
at that period $P$. For a fixed period $P$, the companion mass is
undetectable up to a critical mass, and any higher mass is then
detectable.
The upper mass limits of the two epochs of direct imaging are combined
in a manner similar to Section \ref{sec:dir}, and
the direct imaging limits for elliptical orbits is shown in Panel (b) in
Fig. \ref{fig:ellipt}.
These mass limits are higher than those seen for circular orbits because
we are assuming that these elliptical orbits are oriented with the
semi-major axis pointing along the line of the sight to J1407.
The projected angular separation of J1407b from the primary star is
smaller than the separation for a circular orbit of the same orbital
period, resulting in higher mass limits.

\subsubsection{Limits from Photometric Monitoring for elliptical orbits}

Photometric monitoring rules out any elliptical orbits with $e<0.3$, and
significantly limits orbital eccentricities with $e<0.6$, as periods
shorter than 850 d are ruled completely out.
Longer orbital periods therefore need more eccentric orbits to raise the
periastron velocity to the 32km.s$^{-1}$ required.
This already presents a significant challenge as to how coherent
dynamically long lived structures can exist around a secondary
companion.

\subsubsection{Limits from RV Measurements for elliptical orbits}

We construct an RV
model $f(P,M,e)$ for J1407b assuming periastron at the time of the 2007
transit, resulting in elliptical orbits with their semi-major axis
pointing towards the earth.
We carry out the same procedure as detailed in Section \ref{rvdir} for
determining the fit for these elliptical orbits and the resultant
$\chi^2$ is converted to a probability as seen in Fig.
\ref{fig:ellipt} (c).

\subsubsection{Limits from the Hill sphere for elliptical orbits}

For elliptical orbits, the Hill radius will change with time, reaching a
minimum at periastron.
Given the duration of the eclipse of 54 d, we calculate the size of
the ring system at periastron in units of the Hill sphere $\xi_{peri}$, and this is
shown as the horizontal contours in panel (a) of Figure
\ref{fig:ellipt}.
Periastron passage only comprises a small fraction of the total orbital
period for very eccentric orbits, and the secondary companion will spend
the majority of the orbital period at much larger separations.
We therefore calculate the mean separation $\bar{a}$ which represents the
separation averaged over one orbital period:

\[
\bar{a} = a\left ( 1+\frac{e^2}{2}\right ).
\]

The diameter of the ring system is calculated from $v_{peri}$ and
54 d of eclipse, resulting in a larger ring diameter than for the
equivalent circular orbit with period $P$.
The diameter of ring system in units of the Hill sphere is shown in panel (d) of
Fig. \ref{fig:ellipt}.

\subsubsection{Combined mass-period limits for J1407b for elliptical orbits}

\begin{figure*}
\centering
\includegraphics[angle=0,width=\textwidth]{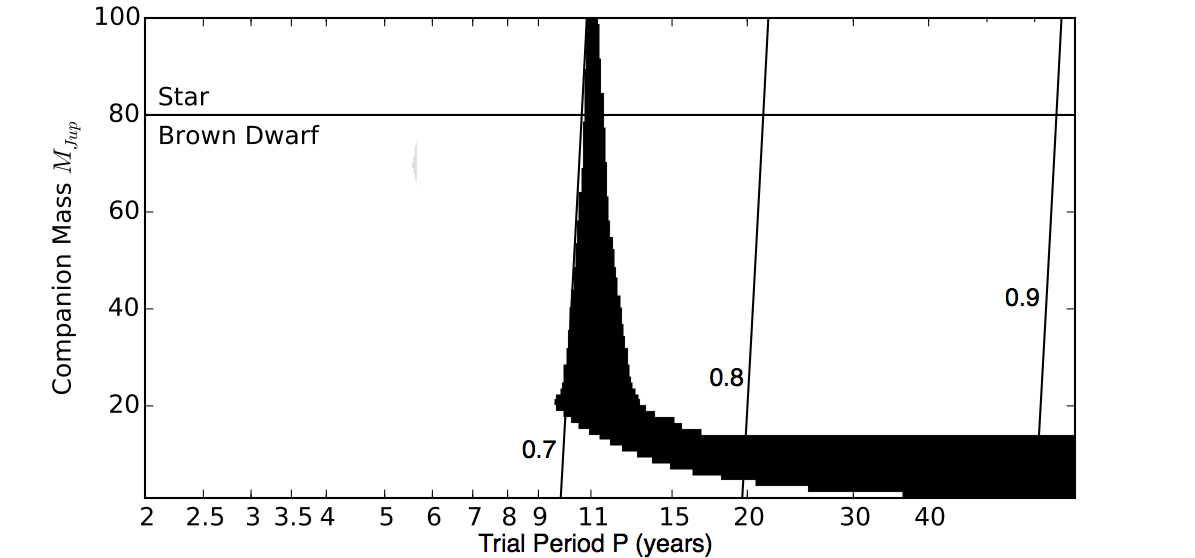}

\caption{Combined mass-period limits for J1407b with elliptical orbits. This figure is a
  logical combination of the panels (b), (c) and (d) in Figure
  \ref{fig:ellipt}. Vertical lines represent ellipticities of 0.7, 0.8
and 0.9 from left to right respectively.
\label{fig:pemassfinalell}}

\end{figure*}

Combining the four sets of constraints listed previously, we show our
limits on the possible mass and orbital period for elliptical orbits of
J1407b in Fig.  \ref{fig:pemassfinalell}. Possible periods and masses are indicated by
shaded regions in the figure. Overplotted are different
values for $e$.

The mass limits for the elliptical orbits are higher than for the
circular orbits. This is a result of choosing elliptical orbits with the
constraint of having periastron occur at the time of the eclipse in 2007
-- orbits of a given period $P$ have smaller semiminor axes compared to
circular orbits, and so the projected angular separation is
correspondingly smaller.
This also increases the mass limits derived from the RV measurements.
After passage through periastron, the secondary companion spends most of
its orbit at distances larger than the semimajor axis, with correspondingly much smaller radial
velocities seen in the star, and much smaller corresponding
accelerations.

A minimum eccentricity of 0.5 is required to keep the ring system within
the Hill sphere, although this is dependent on the precise nature of the
interaction of the secondary on such an eccentric orbit with the
surrounding ring system.
Instead of limits due to gradient velocities (as considered for circular
orbits), we are now limited by the eccentricity of the orbit of J1407b. Table \ref{tab:probee} shows the relative probabilities
for the eccentric orbital solutions, computed in a similar manner to
Table \ref{tab:prob}.
The most significant result is that J1407b requires a highly
eccentric $(e>0.7)$ orbit. This is seen in other young stellar systems
with low-mass companions --
PZ Tel B is a brown dwarf mass companion with eccentricity greater than
0.6 \citep{Biller10,Mugrauer12}.
A more critical constraint is that the ring system will overflow the
Hill radius during the periastron passage, potentially highly perturbing
the system and ring structure seen in the J1407 light curve.
An analysis of the effects of a periastron passage on a giant ring
system is beyond the current scope of this paper.

\begin{table*}
 \centering
 \begin{minipage}{140mm}
    \caption{Probable mass of J1407b for elliptical orbits\label{tab:probee}}
    \begin{tabular}{lcccc}
    \hline
Mass range & $e<0.9$ & $e<0.8$ \\
\hline
`star'        $P(M>80\mj)$  & 0.5\% & 1.2\%  \\
`brown dwarf' $P(13-80\mj)$ & 34\%   & 71\%  \\
`planet'      $P(<13\mj)$   & 65\%   & 28\%  \\
\hline
Probable period $\overline{P}$ (yr)      & 27.5   & 13.3  \\
Probable mass   $\overline{M}$ (\mjup)   & 14.0   & 23.8  \\
\hline
\end{tabular}
\end{minipage}
\end{table*}

% validity of large e orbits and the Universe completely messing with us

\section{Conclusions}
\label{co}

The observations reported here explore the highest angular resolutions
attainable on single dish optical telescopes, along with contrasts that
are not attainable with classical image subtraction techniques at small
inner working angles.
We can place strong constraints on the nature of the companion detected
in transit in \citet{Mamajek12} for circular orbits.
Additional constraints based on high spectral resolution and precision
RV measurements, combined with orbital periods ruled out by
photometric monitoring and dynamical arguments show that J1407b is a
substellar object and likely to be a planetary mass or brown
dwarf mass object.
Stability arguments favour longer periods where $\xi < 0.4$, whilst
observations of rapid changes in the light curve over the course of a
few hours \citep{vanWerkhoven14} strongly suggest that the orbital
period must be short ($P \ltsim\, 15$ yr).
The ring system is also considerably larger than the Roche radius for
the secondary companion, suggesting that we are seeing this ring system
in a transitional state where exosatellites are in the process of
formation.

Using elliptical orbital solutions to provide a large transverse
velocity at periastron results in high orbital eccentricity solutions
$(e>0.7)$, a large ring system that fills a significant fraction of the Hill sphere $(\xi
> 0.5)$ and an overflowing Hill sphere at periastron passage.
These elliptical orbits provide significant challenges for ring structural
stability, either requiring rapid dynamical settling within one orbital
period or the stabilizing influence of exosatellites within the ring
system.

All models imply a ring system for J1407b that fills at least $0.15$ of
the Hill sphere and is significantly larger than the Roche radius.
These large rings are unstable on long time-scales and will
ultimately accrete to form exosatellites, contrasting
with Solar system models where moons are formed from gradual overflow
over the Roche radius \citep{Crida12}.

For future observations, a detection of another eclipse is the most
direct determination of the orbital period of J1407b.
The observations presented in this paper will then give an upper and
lower limit for the mass of the secondary companion and immediately
constrain the nature of the system.
Further RV observations will start to significantly rule out the largest
masses at decade periods. An additional observation using SAM will also
reduce the mass peaks at 5.5 and 11 year periods.

Observations at longer wavelengths may be able to detect and possibly
resolve the ring system around J1407b.
Heating from the secondary companion combined with the large surface
area of the ring system results in significant flux at submm
wavelengths, detectable with the Atacama Large Millimeter Array (ALMA)
or a large single dish submm telescope. 
When the full ALMA suite of
telescopes is deployed, a high precision astrometric detection can
constrain the orbit of the secondary companion and measure
the mass of the ring system.

Additional searches are now being carried out in archival photometric
data to look for similar events that may have been overlooked or
rejected by exoplanetary transit pipelines \citep[e.g.\ ][]{Quillen14}.
The ringed system around J1407b represents a unique laboratory to both
spatially and spectrally resolve a young 16 Myr old disk around a likely
substellar object and probe its structure to unprecedented spatial
scales.
Photometric monitoring is underway to look for the beginning of the next
eclipse, and will signal the start of intensive observational campaigns
over the subsequent weeks.

\section*{Acknowledgements}

% VLT STAFF
We are grateful to the VLT staff and to Nuria Huelamo who was vital in
preparing these observations.
% PROMPT AND APASS
Photometry from the PROMPT-4 telescope and the APASS was made possible
by funding from the Robert Martin Ayers Science Fund.
% ADS & SKYVIEW
This research has made use of NASA's Astrophysics Data System and
Skyview.
% SIMBAD & CDS
This research has made use of the SIMBAD database, operated at CDS,
Strasbourg, France.
% VLT
Some of the observations were taken at the VLT (program identifier
090.C-0827(A)).
% TRIAUD
AHMJT received funding from the Swiss National Science
Foundation in the form of an Advanced Mobility Post-doctoral Fellowship
(P300P2-147773).
% LACOUR
SL acknowledges support by the French National Agency for Research
(ANR-13-JS05-0005-01).
% EEM 
EEM acknowledges support from NSF grants AST-1008908 and AST-1313029.
% VOSA
This publication makes use of VOSA, developed under the Spanish Virtual
Observatory project supported from the Spanish MICINN through grant
AyA2011-24052.
% CORALIE
The Swiss Euler Telescope is operated by the University of Geneva
and is funded by the Swiss National Science Foundation.  We thank the
many observers that obtained CORALIE data on this target and appreciate
the technical assistance that was provided by the Observatory of Geneva.
We also thank the kind attention of the ESO staff at La Silla
observatory.
% KECK
Some of the data presented herein were obtained at the W.M. Keck
Observatory, which is operated as a scientific partnership among the
California Institute of Technology, the University of California and the
National Aeronautics and Space Administration.
The Observatory was made possible by the generous financial support of
the W.M. Keck Foundation.
The authors wish to recognize and acknowledge the very significant
cultural role and reverence that the summit of Mauna Kea has always had
within the indigenous Hawaiian community.  We are most fortunate to have
the opportunity to conduct observations from this mountain.
% WISE
AllWISE makes use of data from WISE, which is a joint project of the
University of California, Los Angeles, and the Jet Propulsion
Laboratory/California Institute of Technology, and NEOWISE, which is a
project of the Jet Propulsion Laboratory/California Institute of
Technology. WISE and NEOWISE are funded by the National Aeronautics and
Space Administration.
% REF
We thank the anonymous referee for their comments and suggestions on
this paper.

\bibliographystyle{mn}
\bibliography{mn-jour,kenworthy}

\label{lastpage}
\end{document}